\documentclass[aps,prb,reprint,groupedaddress]{revtex4-2}

\usepackage{amsmath,amssymb}
\usepackage{bm}
\usepackage{graphicx}
\usepackage{xcolor}
\usepackage{hyperref}
\usepackage[normalem]{ulem}

\begin{document}

\title{World Models for Learning Quantum Many-Body Dynamics}

\author{Yuan Wang}
\affiliation{School of Mathematical and Physical Sciences, University of Sheffield, Sheffield S10 2TN, United Kingdom}

\author{Oleksandr Kyriienko}
\affiliation{School of Mathematical and Physical Sciences, University of Sheffield, Sheffield S10 2TN, United Kingdom}

\begin{abstract}
We propose using world models for learning Hamiltonians from measured dynamics of a quantum computer or simulator. Implicitly, the parameters of a Hamiltonian are encoded in how the measured few-body observables evolve. By using a joint-embedding predictive architecture (JEPA) to perform prediction in latent space, we show that the device Hamiltonian can be efficiently inferred from noisy measurements. We train a spatiotemporal encoder and predictor with approximately $2.7$ million trainable weights using a single objective, to predict the representation of the next window of measurements from the previous one. The network is thereby driven to retain the reproducible dynamics and discard shot noise. We recover the field and coupling maps of a two-dimensional Ising lattice in real and imaginary time, and the disorder map of a one-dimensional Heisenberg chain of $100$ sites. In every settings the encoder achieves lower error than a masked autoencoder and than the same encoder left untrained. It also outperforms the same regression head applied directly to time-averaged observables, by factors of $4.4$ to $30$. We establish latent prediction as a practical self-supervised objective for learning from measurements of quantum dynamics.
\end{abstract}

\maketitle

\section{Introduction}
\label{sec:intro}

The cost of simulating many-body quantum dynamics grows exponentially with system size~\cite{altman2021quantum,daley2022practical}. Several quantum platforms have reached sizes that are challenging for exact classical simulation, including Rydberg atom arrays~\cite{bernien2017probing,ebadi2021quantum,scholl2021quantum,Mendessantos2026}, trapped ions~\cite{zhang2017observation}, and superconducting circuits~\cite{arute2019quantum,kim2023evidence}. 
Current experiments probe particular observables and evolution times at which approximate classical methods also struggle. On a $103$-qubit superconducting processor, second-order out-of-time-order correlators were measured at scales that are not known to be tractable by classical algorithms~\cite{google2025constructive,bermejo2026tnbp}. A $56$-qubit trapped-ion processor simulated the dynamics of a two-dimensional (2D) Fermi-Hubbard model, finding deviations from tensor-network (TN) predictions~\cite{alam2025fermionic}. 
Digitized Ising dynamics also showed thermalization on timescales challenging for classical simulation~\cite{haghshenas2026digital}. 

Without a classical reference, verifying the accuracy of computed observables becomes a challenge~\cite{barron2026observable}. Quantum simulation experiments typically probe the dynamics of a Hamiltonian specified in advance. A separate question is which Hamiltonian the device implements. The nominal controls set the intended fields and couplings, but spatial inhomogeneity, cross-talk and drift can shift the realized parameters~\cite{browaeys2020many,gross2017quantum}. Even a precisely specified target therefore leaves a certification question, particularly at sizes where classical computation can no longer check the answer~\cite{eisert2020quantum}. Full state tomography is exponentially costly in general~\cite{haah2016sample}, although efficient protocols exist for structured states~\cite{cramer2010efficient}. Randomized measurements and classical shadows give access to few-body observables~\cite{huang2020predicting,elben2023randomized}. Targeted measurements can also characterize a device~\cite{da2011practical} or reveal the entanglement structure of its state~\cite{kokail2021entanglement}. 

Hamiltonian learning asks which generator produced the observed dynamics~\cite{gebhart2023learning}. Its parameters can be inferred from stationary states~\cite{bairey2019learning,qi2019determining,valenti2019hamiltonian}, including Gibbs states~\cite{anshu2021sample}, or from time-resolved measurements~\cite{granade2012robust,wiebe2014hamiltonian,zubida2021optimal,hangleiter2024robustly,stilck2024efficient,wilde2026scalably}. Formal results now cover learning local Hamiltonians from Gibbs states in polynomial time at a fixed temperature~\cite{bakshi2024learning}, learning many-body Hamiltonians at the Heisenberg limit under controlled evolution~\cite{huang2023learning}, and learning without an assumed interaction structure~\cite{hu2025ansatz}. Experiments have used a photonic simulator to learn the Hamiltonian of a solid-state spin~\cite{wang2017experimental} and inferred models of quantum systems directly from experimental data~\cite{gentile2021learning}. TN emulation tested Hamiltonian learning on spin chains of several hundred sites~\cite{johnson2026hamiltonian}. These estimators are tied to a fixed measurement protocol or model structure, and have to be redesigned when the parameterization or the readout changes.
\begin{figure*}[t]
\centering
\includegraphics[width=\textwidth]{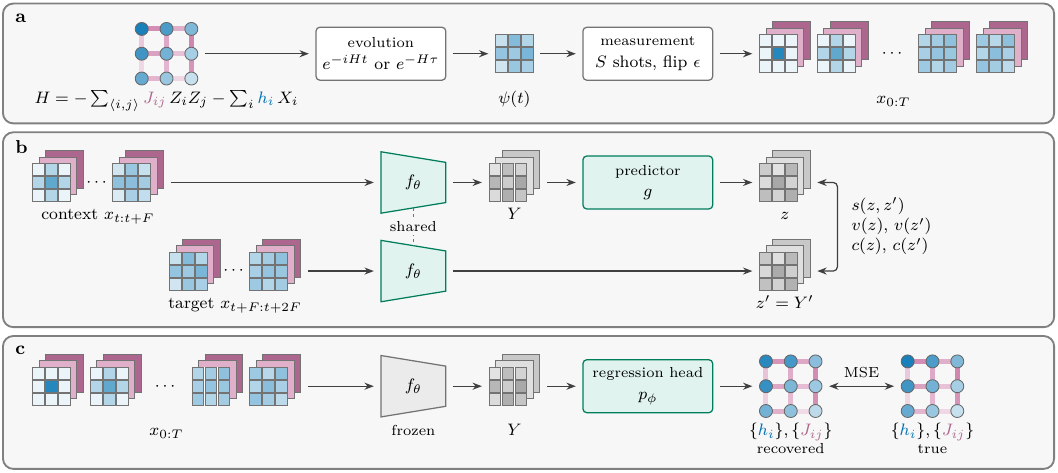}
\caption{\label{fig:pipeline}
Overview of the framework. (a) The two-dimensional transverse-field Ising model, with a field $h_i$ on every site (blue) and a coupling $J_{ij}$ on every bond (purple), is evolved in real or imaginary time and its state $\psi(t)$ is read out by noisy single-shot measurements ($S$ shots with a detection flip of probability $\epsilon$, Fig.~\ref{fig:measurement}), yielding an observable trajectory $x_{0:T}$ in the form of a multichannel spatiotemporal field. The disordered Heisenberg chain enters the same pipeline, presented as a one-dimensional strip of sites. (b) Joint-embedding predictive architecture (JEPA) pretraining, in which the encoder-predictor pair acts as a latent world model of the trajectory: a shared encoder $f_\theta$ embeds a context window $x_{t:t+F}$ of $F$ frames and the target window that follows it into the representations $Y$ and $Y'$, and a predictor $g$ maps $Y$ to an embedding $z$, which the training objective [Eq.~\eqref{eq:vicreg}] draws toward the target embedding $z'=Y'$. (c) Parameter recovery: the pretrained encoder is frozen and a lightweight regression head $p_\phi$ is trained to recover the governing map from the representation $Y$ of the trajectory, by the mean-squared error (MSE) against the true map. The target is the field $\{h_i\}$, or the field and the couplings $\{J_{ij}\}$ together. Networks are drawn in green while they are trained and in grey once frozen.}
\end{figure*}

Machine learning (ML) has changed what can be extracted from physical data~\cite{carleo2019machine,dawid2022modern,carrasquilla2017machine,van2017learning,rem2019identifying,bohrdt2019classifying,carleo2017solving,torlai2018neural}. Large models pretrained on many systems improve prediction and sample efficiency on new ones. A model pretrained on families of partial differential equations transfers to equations it has not seen with far fewer examples than training from scratch~\cite{herde2024poseidon}. Beyond predicting outcomes, learning can identify the laws that generate them. Sparse regression over a library of candidate terms recovers governing equations from time series~\cite{brunton2016discovering}, and physics-informed networks constrain the fit by the equations themselves~\cite{raissi2019physics}. In the quantum setting, differentiable quantum circuits have been used to discover the differential equation behind observed data~\cite{heim2021quantum}, and quantum Fokker-Planck solvers trained as generative models infer the parameters of a stochastic process from samples of its distributions~\cite{mahassen2026quantum}. Generalizing from limited quantum data is nevertheless hard. For the learning of unitary dynamics, out-of-distribution generalization has been established under stated conditions on the training set~\cite{caro2023out}, which shows what a learned model of quantum dynamics can support once those conditions are met.

Much of this progress in ML for physics concerns learning operators and predicting trajectories. Neural operators map initial conditions or parameter fields to solutions of differential equations~\cite{azizzadenesheli2024neural}, and pretraining across physical systems can improve predictions on new ones~\cite{mccabe2024multiple,subramanian2023towards}. Self-supervised methods learn representations from different views of the same data~\cite{chen2020simple,zbontar2021barlow}. For physical systems, such views can be generated by applying symmetries to solutions of differential equations~\cite{mialon2023self}. A \emph{world model} learns an internal representation of a system from observations and predicts how it evolves~\cite{ha2018world,lecun2022path,hafner2025mastering}. In a joint-embedding architecture, the prediction target is the representation of a future observation~\cite{lecun2022path,assran2025v}. We ask whether predicting the representation of the next measurement window preserves enough information to identify the Hamiltonian of a quantum system.

We propose a latent world model of measured quantum dynamics based on the joint-embedding predictive architecture (JEPA)~\cite{lecun2022path,assran2023self,bardes2024revisiting}. An encoder sees a short window of measurement frames, and a predictor forecasts the representation of the window that follows. The same Hamiltonian governs both windows, whereas every frame carries fresh shot noise. Hence, a model trained to predict the next window gains nothing from retaining the noise and is driven to retain the reproducible dynamics. Pretraining with this objective needs no parameter labels, which matches the regime that hardware produces, where unlabelled trajectories are abundant and labelled ones come only from classically tractable simulation. We therefore pretrain the encoder without labels and fit a small labelled regression head that reads out the fields and couplings~\cite{balestriero2023cookbook}. On a 2D transverse-field Ising lattice in real and imaginary time, and on a disordered Heisenberg chain of $100$ sites, latent prediction recovers the parameter maps with lower error than a masked autoencoder (MAE)~\cite{he2022masked,tong2022videomae,feichtenhofer2022masked} trained on the same data and than the same encoder left untrained. The recovered Hamiltonian also forecasts the dynamics, predicting the remaining frames of a record from its first quarter close to the shot-noise limit.

\section{Physical system}
\label{sec:system}

\subsection{System Hamiltonians}
\label{sec:hamiltonian}

We begin by formulating a learning problem for two classes of spin-$1/2$ lattice Hamiltonians with spatially varying parameters, which serve complementary purposes. The 2D transverse-field Ising model (TFIM) provides a controlled testbed where the quantum dynamics can be simulated exactly, allowing us to study the recovery of local fields and couplings, both separately and jointly, under real- and imaginary-time evolution. The one-dimensional (1D) disordered Heisenberg chain serves as a test of scalability to larger systems. Although its dynamics can be simulated using TN methods, direct simulation in the full $2^{100}$-dimensional Hilbert space is computationally prohibitive. This motivates the development of world models that learn directly from measured dynamics, with the longer-term aim of addressing systems where classical simulation becomes impractical but quantum hardware can provide sufficient measurement data.

\emph{Transverse-field Ising model.} Our first testbed is based on a 2D $L_x\times L_y$ square lattice with periodic boundaries,
\begin{equation}
H = -\sum_{\langle i,j\rangle} J_{ij}\, Z_i Z_j - \sum_i h_i\, X_i,
\label{eq:tfim}
\end{equation}
where $X_i$, $Z_i$ are Pauli operators on site $i$, $J_{ij}$ is the nearest-neighbour coupling on the bond $\langle i,j\rangle$, and $h_i$ is a site-dependent transverse field. We prepare the lattice in a fully polarized product state with a single flipped spin at the centre, and evolve it under $H$. The localized excitation spreads at a rate set by $\{J_{ij}\}$ and precesses at a rate set by $\{h_i\}$, and the governing map determines the time-dependent observables. In Fig.~\ref{fig:pipeline} we show the resulting pipeline. We study the dynamics of this model rather than its equilibrium phases, and specifically the inverse problem of inferring from an observed evolution a governing parameter map: the inhomogeneous field $\{h_i\}$, alone or jointly with the coupling map $\{J_{ij}\}$. Each trajectory draws an independent smooth parameter map from a low-dimensional ensemble (Appendix~\ref{app:fields}), so the target is a distinct 2D surface for every run, and its recovery requires a representation that preserves nontrivial spatial structure rather than a single scalar. The spatially correlated ensemble follows the physical origin of the inhomogeneity. The dominant inhomogeneities in tweezer arrays and optical lattices originate in the transverse intensity profile of the addressing beams and in slowly varying detunings~\cite{browaeys2020many,gross2017quantum}, which yield parameter maps correlated across the array rather than site-wise random. Simulation parameters and settings are described in Table~\ref{tab:models}.

\emph{Disordered anisotropic Heisenberg chain.} To demonstrate that the framework is not tied to small system size, we also study the disordered anisotropic Heisenberg chain of $n=100$ sites of Ref.~\cite{wilde2026scalably}, with open boundaries,
\begin{equation}
H = \sum_{i=1}^{n-1} \big( J_x X_i X_{i+1} + J_y Y_i Y_{i+1}
      + J_z Z_i Z_{i+1} \big) + \sum_{i=1}^{n} h_i\, X_i.
\label{eq:heis}
\end{equation}
The chain is prepared in the fully polarized product state $|0\rangle^{\otimes n}$ and evolved in real time by time-evolving block decimation (TEBD) on matrix product states~\cite{white1992density,vidal2003efficient,vidal2004efficient,schollwock2011density,paeckel2019time}. The model and its parameters follow Ref.~\cite{wilde2026scalably} and are listed in Table~\ref{tab:models}. The time grid and the bond dimension $\chi$ of the matrix-product state are our distinct choices. This setup differs from Ref.~\cite{wilde2026scalably} in two respects: disorder realizations are resampled per trajectory rather than fixed, and observables use the finite-shot channels described in Sec.~\ref{sec:measurement} instead of randomized Pauli bases. Sec.~\ref{sec:heis} reports the corresponding results.

We focus primarily on real-time (unitary) quench dynamics, though the framework also accommodates imaginary-time evolution, $e^{-H\tau}$. This relaxes the state toward the ground state of $H$, encoding parameters within the relaxation trajectory rather than in a coherent quench. Imaginary-time evolution is applied exclusively to the Ising model, whereas the Heisenberg chain is evolved strictly in real time.
\begin{table*}[t!]
\caption{\label{tab:models}
The two Hamiltonians. The Ising model is the controlled testbed, exact for every task variant. The Heisenberg chain tests the same pipeline where exact simulation is not available. i.i.d.: independent and identically distributed.}
\begin{ruledtabular}
\begin{tabular}{lll}
 & Transverse-field Ising [Eq.~\eqref{eq:tfim}] & Disordered Heisenberg chain [Eq.~\eqref{eq:heis}] \\
\colrule
Geometry            & $4\times3$ square lattice, periodic        & open chain, $n=100$ sites \\
System size         & $12$ qubits (dimension $4096$)             & $100$ qubits \\
Recovery target     & $\{h_i\}$, alone or jointly with $\{J_{ij}\}$   & on-site disorder field $\{h_i\}$ \\
Parameter ensemble  & smooth low-frequency surface, per trajectory & i.i.d.\ $\mathcal{U}[-1,1]$ per site, per trajectory \\
Fixed parameters    & the parameter that is not the target        & $(J_x,J_y,J_z)=(-1.0,-0.5,-0.4)$ \\
Initial state       & polarized, one flipped spin at the centre   & $|0\rangle^{\otimes n}$ \\
Dynamics            & real time (imaginary time also available)   & real time \\
Propagation         & exact sparse statevector                    & TEBD on matrix product states, $\chi=128$ \\
Readout channels    & $\langle n_i\rangle$ with bond $\langle n_i n_j\rangle$ & $\langle X_i\rangle,\langle Y_i\rangle,\langle Z_i\rangle$ and $\langle Z_i Z_{i+1}\rangle$ \\
Trajectory          & $T=32$ frames, spacing $\delta t = 0.15$ ($\delta\tau=0.1$ in imaginary time)   & $T=32$ frames, spacing $\Delta t = 0.1$ \\
Channel count $C$  & $3$      & $4$ \\
Encoder window $F$ & $8$ frames                            & $16$ frames (Section~\ref{sec:heis}) \\
\end{tabular}
\end{ruledtabular}
\end{table*}

\subsection{Measurement}
\label{sec:measurement}
\begin{figure}[t!]
\centering
\includegraphics[width=\columnwidth]{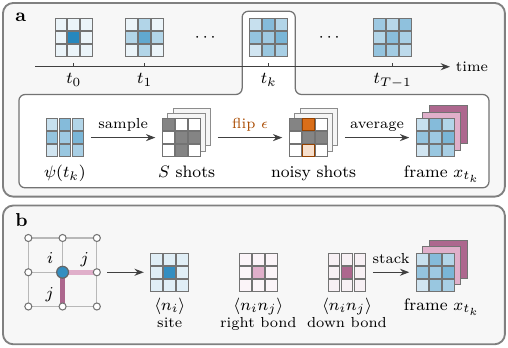}
\caption{\label{fig:measurement}%
Measurement model. (a) The state is evolved on a uniform time grid, and each frame is measured independently: $S$ single shots of the \emph{same} state $\psi(t_k)$ are drawn by Born-rule sampling in the computational basis, a symmetric detection flip of probability $\epsilon$ corrupts every shot, and averaging over shots yields the estimates $\langle n_i\rangle$ and $\langle n_i n_j\rangle$ that constitute the frame $x_{t_k}$. (b) Single frame: the site occupations and the right- and down-bond correlators (each placed at the bond's first site) enter as co-located channels of an $L_x\times L_y$ multichannel image. For the Heisenberg chain the frame is a one-dimensional strip of $n=100$ sites with the four channels $\langle X_i\rangle,\langle Y_i\rangle,\langle Z_i\rangle,\langle Z_iZ_{i+1}\rangle$.}
\end{figure}

The lattice is measured in the occupation basis, as illustrated in Fig.~\ref{fig:measurement}(a). This output, native to analog platforms~\cite{bakr2009quantum,sherson2010single,gross2021quantum}, yields site-resolved bitstrings from which commuting occupations $\langle n_i\rangle$ (with $n_i=(1-Z_i)/2$) and bond correlators $\langle n_i n_j\rangle$ are extracted simultaneously.

We summarize the state across $T$ time steps using a small set of few-body observables accessible from single-shot measurements. We include the measurement noise explicitly: each observable is estimated from $S$ shots per frame, subject to an optional symmetric per-site detection flip with probability $\epsilon$. This error modifies estimators as $\langle n_i\rangle\to(1-2\epsilon)\langle n_i\rangle+\epsilon$ and $\langle Z_iZ_j\rangle\to(1-2\epsilon)^2\langle Z_iZ_j\rangle$, contracting an $m$-body observable toward its unpolarized value by $(1-2\epsilon)^m$. Consequently, shot noise and readout bias enter as unbiased variance and a deterministic contraction factor.

The lattice readout is sampled directly from full statevectors. For the Heisenberg chain, tensor-network propagation yields exact expectation values of $\langle X_i\rangle$, $\langle Y_i\rangle$, $\langle Z_i\rangle$ and $\langle Z_iZ_{i+1}\rangle$, to which the finite-measurement noise model is applied independently per observable (Appendix~\ref{app:procedure}). Consequently, the chain's channel noise is uncorrelated, unlike the occupation readout, which inherits correlations from shared bitstrings.

In both systems the per-frame observables form a multichannel lattice image [Fig.~\ref{fig:measurement}(b)], with $C=3$ channels for the lattice and $C=4$ for the chain. A trajectory of $T$ frames thus constitutes a multichannel spatiotemporal field with the same tensor structure as a video. Working at the level of observables avoids the exponential cost of state tomography while retaining the dynamics, so that architectures built for video apply directly.

\section{Self-supervised learning of a latent world model}
\label{sec:methods}

\subsection{Two pretraining schemes with a shared encoder}
\label{sec:objectives}

The question is whether self-supervised representations of measured dynamics encode the governing parameters, and which training objective extracts them more accurately. Single-window observables form a noisy, high-dimensional encoding of a state governed by few physical parameters. A useful representation compresses this signal to those few parameters while filtering the shot noise. We compare two self-supervised schemes sharing an identical spatiotemporal convolutional encoder $f$ operating over $F$-frame windows ($F=8$, except in Sec.~\ref{sec:heis}). Because these schemes differ solely in their training objectives, performance variations reflect the training objectives rather than model capacity (Appendix~\ref{app:arch}).

The JEPA~\cite{lecun2022path,assran2023self,bardes2024revisiting} [Fig.~\ref{fig:pipeline}(b)] predicts, in the learned latent space, the representation of a future window of the trajectory from a past one: a predictor $g$ maps the context embedding $f(\mathrm{context})$ toward the target embedding $f(\mathrm{target})$ by minimizing the VICReg cost function~\cite{bardes2021vicreg},
\begin{equation}
\mathcal{L} \;=\; \lambda\, s(z,z') \;+\; \mu\,\big[v(z)+v(z')\big]
\;+\; \nu\,\big[c(z)+c(z')\big],
\label{eq:vicreg}
\end{equation}
with $z=g(f(\mathrm{context}))$ and $z'=f(\mathrm{target})$. The invariance term $s$ is the mean-squared distance between $z$ and $z'$ and aligns prediction with target, the variance term $v$ penalizes any embedding dimension whose standard deviation falls below unity, and the covariance term $c$ drives the off-diagonal covariances of the embedding toward zero. The last two prevent the collapse that a purely predictive cost function would otherwise produce, in which every window is mapped to the same embedding. Eq.~\eqref{eq:vicreg} does not reference the observable values. In Appendix~\ref{app:arch} we provide the coefficients, the space-time positions over which the terms are evaluated, and the stride at which windows are drawn.

The encoder-predictor pair $(f,g)$ forms a world model of the measured quantum trajectory~\cite{ha2018world,lecun2022path,assran2025v}. The spatiotemporal encoder $f$ maps a measurement window into a latent state, while the predictor $g$ advances this representation to subsequent windows entirely within latent space. Within a window, $f$ captures the temporal structure directly, through three-dimensional (3D) convolutions over space and time. The target representation $z'=f(\text{target})$ is produced by the same encoder, rather than by a decoder, such that no individual measurement outcome is reconstructed. The regression heads of Sec.~\ref{sec:probe} read the encoder alone, and they quantify the parameter information retained within the frozen latent representation $f$. In Sec.~\ref{sec:dynamics} we return to the predictor, roll it forward from the first window of a record to forecast the rest without any Hamiltonian, and compare that forecast with the one produced by the recovered Hamiltonian.

The approach based on MAE~\cite{he2022masked,tong2022videomae,feichtenhofer2022masked} [Fig.~\ref{fig:pipeline_mae_raw}(a)] instead reconstructs the observables at lattice sites concealed by a space-time tube mask, which hides the same sites in every frame of the window, minimizing the reconstruction error directly in observable space. The mask ratio and the sweep behind it are given in Appendix~\ref{app:fairness}. The two objectives therefore differ in what they retain. The MAE loss decreases when the encoder keeps the microscopic detail of the measured observables, including shot noise, whereas the JEPA loss decreases only when it keeps the component of one window that predicts the next, which is the component determined by the Hamiltonian.

\subsection{Evaluating the frozen representation}
\label{sec:probe}

To quantify the physical content of a representation, we freeze the pretrained encoder and train a lightweight regression head to recover the parameter map [Fig.~\ref{fig:pipeline}(c)]. The regression head temporally averages feature maps across all $T/F$ non-overlapping trajectory windows, applies three $1\times 1$ convolutions, and resizes the output to the lattice geometry, recombining channels locally without spatial mixing across sites. Target outputs depend on the task: a single-channel field map $\{h_i\}$ ($L_x\times L_y$), a stacked three-channel map of the field and the horizontal and vertical couplings $\{J_{ij}\}$ for joint recovery, or an $n$-component 1D map $\{h_i\}$ for the Heisenberg chain. Because the head cannot transport information between sites, any recovered spatial structure must originate from the frozen representation rather than from the capacity of the head.

Recovery performance is measured using standardized mean-squared error (MSE), where an MSE of unity corresponds to predicting the target map mean. Targets are standardized per channel to balance field and coupling parameters in both the training and the evaluation of the head. For joint tasks, errors are reported both in aggregate and split by parameter type.

As a control, we evaluate the same regression head on time-resolved and time-averaged observables without a learned encoder [Fig.~\ref{fig:pipeline_mae_raw}(b, c)]. Any performance gap over the time-averaged control therefore measures dynamical information rather than a difference in the observables supplied. The head architecture, its training protocol and its tuning grid (Appendix~\ref{app:fairness}) are the same for every representation, so the rows of each table differ only in the input representation. Global setting variations across tables are detailed in Appendix~\ref{app:config}, and the complete protocol is provided in Appendix~\ref{app:procedure}.

\section{Results}
\label{sec:results}

We proceed to describe results of training world models and quantify the accuracy of learning model parameters. 
Unless stated otherwise, results are reported for the $4\times 3$ lattice of Table~\ref{tab:models}, with $S=512$ shots per frame and a detection error rate of $\epsilon=0.02$. Each dataset comprises $700$ training, $100$ validation, and $200$ test trajectories, shared identically between the self-supervised and head-training stages. Quoted values are the mean $\pm$ one sample standard deviation across four independent pipeline seeds. Since datasets are generated once and shared bit-identically across all models, this spread reflects the optimization alone, not the sampling of the data. Hyperparameters and configuration details are provided in Appendix~\ref{app:config} and the selection of the learning rates in Appendix~\ref{app:fairness}.
\begin{table*}[t!]
\caption{\label{tab:joint}
Joint recovery error of field and couplings in standardized MSE, lower is better. The last three columns repeat the task on the same Hamiltonians evolved in imaginary time (Section~\ref{sec:imag}). The first two rows are the frozen pretrained encoders, and the other three involve no training of the encoder. Learning rates and the run configuration are listed in Appendix~\ref{app:config}. The most accurate input to the regression head is in bold.}
\begin{ruledtabular}
\footnotesize\medmuskip=0mu 
\begin{tabular}{lcccccc}
Representation & \shortstack{Field $\{h_i\}$\\(real time)} & \shortstack{Couplings $\{J_{ij}\}$\\(real time)} & \shortstack{Overall\\(real time)} & \shortstack{Field $\{h_i\}$\\(imag.\ time)} & \shortstack{Couplings $\{J_{ij}\}$\\(imag.\ time)} & \shortstack{Overall\\(imag.\ time)} \\
\colrule
Latent prediction (JEPA) & $0.0389 \pm 0.0024$ & $\bm{0.1002 \pm 0.0009}$ & $\bm{0.0798 \pm 0.0010}$ & $\bm{0.0585 \pm 0.0013}$ & $\bm{0.2028 \pm 0.0056}$ & $\bm{0.1547 \pm 0.0041}$ \\
Reconstruction (MAE)     & $0.0460 \pm 0.0070$ & $0.1748 \pm 0.0012$ & $0.1319 \pm 0.0021$ & $0.0842 \pm 0.0029$ & $0.3258 \pm 0.0059$ & $0.2453 \pm 0.0046$ \\
Random encoder & $0.0771 \pm 0.0067$ & $0.2100 \pm 0.0129$ & $0.1657 \pm 0.0095$ & $0.0906 \pm 0.0059$ & $0.3532 \pm 0.0377$ & $0.2656 \pm 0.0269$ \\
Time-resolved observables & $\bm{0.0215 \pm 0.0016}$ & $0.4349 \pm 0.0107$ & $0.2971 \pm 0.0068$ & $0.1191 \pm 0.0030$ & $0.8323 \pm 0.0073$ & $0.5946 \pm 0.0048$ \\
Time-averaged observables &  $0.4557 \pm 0.0012$ &  $0.8242 \pm 0.0022$ &  $0.7013 \pm 0.0019$ & $0.1934 \pm 0.0004$ & $0.9178 \pm 0.0009$ & $0.6763 \pm 0.0006$ \\
\end{tabular}
\end{ruledtabular}
\end{table*}

Each table of this section supports three comparisons. The primary one is between the two objectives, all else being shared (Sec.~\ref{sec:objectives}). The second comparison measures the gap between the learned representations and the time-averaged control, testing whether parameter information survives in a purely static description of the dynamics. The third comparison sets the learned representations against the random encoder and the time-resolved observables, which keep the dynamics without any training, thus separating the contribution of pretraining from that of the time dependence. Sec.~\ref{sec:jointresults} also puts the recovered Hamiltonian to use as a generator of dynamics.

\subsection{Recovering the full Hamiltonian}
\label{sec:jointresults}
\label{sec:dynamics}

The inverse problem introduced in Sec.~\ref{sec:hamiltonian} requires recovering the full generator, comprising both field $\{h_i\}$ and coupling $\{J_{ij}\}$ parameters in Eq.~\eqref{eq:tfim}. We therefore begin with the joint task, in which both parameter sets vary per trajectory and the regression head must return the complete Hamiltonian.

\begin{figure}[t]
\centering
\includegraphics[width=\columnwidth]{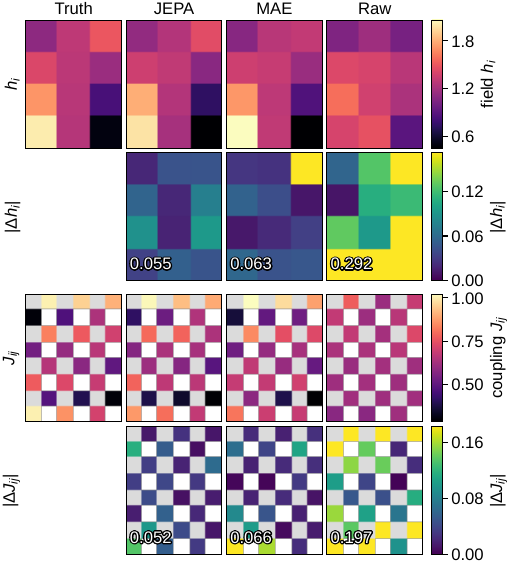}
\caption{\label{fig:maps}%
Recovered field and coupling maps on the joint task of Table~\ref{tab:joint} in real time, for one held-out test trajectory. Columns are the ground truth followed by the regression-head outputs of the latent-prediction (JEPA), reconstruction (MAE) and time-averaged-observable (Raw) representations. The first two rows show the site field, the map $h_i$ and the absolute residual $|\Delta h_i|$ against the truth, the last two the bond couplings, $J_{ij}$ and $|\Delta J_{ij}|$. A coupling panel draws each bond in the cell between the two sites it joins, the sites shaded grey and the plaquette centres blank, so that the position of a cell gives the direction of the bond.  The residual scales are set by the two learned representations, which leaves the raw control saturated. Its magnitude is given by the root-mean-square error inset in each residual panel. All three representations are the seed-0 fits of the runs behind Table~\ref{tab:joint} (Appendix~\ref{app:config}).}
\end{figure}

Table~\ref{tab:joint} reports recovery errors for the joint task. Its first three columns are for real-time evolution and are discussed here, and its last three are for imaginary time (Sec.~\ref{sec:imag}). The two objectives separate clearly: latent prediction achieves an overall error of $0.0798\pm0.0010$, outperforming reconstruction ($0.1319\pm0.0021$) by a factor of $1.7$. The lead comes mainly from the couplings, where the factor is $1.7$ across both bond directions. On the field it is $1.2$, a difference comparable to the seed-to-seed spread of reconstruction. Both learned representations substantially outperform the time-averaged control (by overall factors of $8.8$ and $5.3$, respectively). That gain reflects mostly the time dependence rather than pretraining: the random encoder reaches $0.1657\pm0.0095$, so latent prediction and reconstruction improve on an untrained network by factors of $2.1$ and $1.3$. The time-resolved observables with no encoder reach $0.297\pm0.007$: they recover the field more accurately than either encoder ($0.0215$) but fail on the couplings ($0.43$), which cannot be read from the record of a single site. Every representation yields higher errors on couplings than on fields (by roughly a factor of 2.6 for latent prediction and 1.8 for the control), causing overall errors to be dominated by bond parameters. In Fig.~\ref{fig:maps} we show the recovered maps, the true maps and the residuals for one test trajectory.
\begin{figure}[b!]
\centering
\includegraphics[width=0.9\columnwidth]{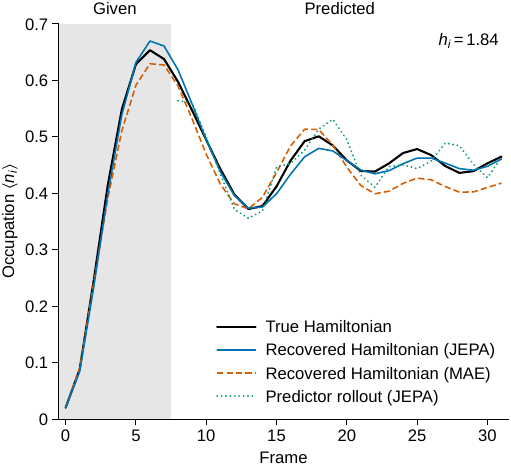}
\caption{\label{fig:forecast}%
Forecast by the recovered Hamiltonian, joint task. The regression heads read only the first $8$ frames of a test trajectory (shaded). The Hamiltonian rebuilt from the field and coupling maps they return is evolved from the initial state, and its occupation at one site is compared over the remaining $24$ frames with that of the true Hamiltonian. The dotted curve is a forecast of the same frames without a Hamiltonian: the predictor of latent prediction rolled forward from the embedding of the shaded frames and read out to the occupations.}
\end{figure}

The recovered maps can also be tested by the dynamics they generate. For every test trajectory we rebuild the Hamiltonian of Eq.~\eqref{eq:tfim} from the maps recovered in real time (Table~\ref{tab:joint}) and evolve the initial state under it. Over the measured record its occupations differ in root mean square from those of the true Hamiltonian by $0.0304\pm0.0005$ for latent prediction and $0.0340\pm0.0008$ for reconstruction, $1.5$ and $1.7$ times the shot noise of a single frame and $3.6$ and $3.2$ times smaller than the error of the ensemble mean of the training trajectories. Recovered from the first $8$ of the $32$ frames alone, the Hamiltonian predicts the remaining $24$ with errors of $0.0261\pm0.0003$ and $0.0317\pm0.0005$, $1.2$ and $1.5$ times their shot noise. In Fig.~\ref{fig:forecast} we show the forecast at one site. The predictor of latent prediction, rolled forward from the same frames without a Hamiltonian, reaches $0.0717\pm0.0091$, $2.7$ times the error of the Hamiltonian recovered through its encoder.

\subsection{A disordered Heisenberg chain of 100 sites}
\label{sec:heis}

Exact statevector propagation is limited to small systems such as the $12$-qubit lattice, but the learning framework itself does not depend on exact solvers. We therefore evaluate the identical pipeline on the $n=100$ disordered anisotropic Heisenberg chain of Eq.~\eqref{eq:heis}, propagated by TEBD (Sec.~\ref{sec:hamiltonian}). The chain enters the pipeline as a 1D array of $100$ sites with the same shot count ($S=512$) and detection error ($\epsilon=0.02$), and the target is the $100$-component disorder map $\{h_i\}$. To capture the temporal signature of i.i.d. disorder, we extend the encoder window to $F=16$ frames (compared to $F=8$ on the lattice). The window, the batch size and an early stop of the fit of the head (Appendix~\ref{app:arch}) are the only protocol adjustments in this work, and they apply to every row of Table~\ref{tab:heis}.

In Table~\ref{tab:heis} we report the results of this benchmark, which lacks smooth spatial correlations and contains as many governing parameters as sites. Latent prediction achieves an error of $0.0020\pm0.0001$ compared to $0.0027\pm0.0000$ for reconstruction (a $1.3$-fold advantage), while the time-averaged control reaches $0.0416\pm0.0006$. The random encoder reaches $0.0040\pm0.0003$, $2.0$ times the error of latent prediction. The time-resolved observables with no encoder reach $0.0024\pm0.0000$: as on the lattice, a single-site field can largely be read from the record of the site itself, and only latent prediction improves on that record, by a factor of $1.2$. The advantage of latent prediction thus carries over to a different Hamiltonian, spatial dimension, propagation algorithm and target distribution. Since the readout differs (Sec.~\ref{sec:measurement}), the absolute errors are comparable within Table~\ref{tab:heis} but not with those of the lattice.
\begin{table}[h]
\caption{\label{tab:heis}
Disorder-field recovery error for the $n=100$ Heisenberg chain propagated by TEBD. Best in bold.}
\begin{ruledtabular}
\begin{tabular}{lc}
Representation & Recovery error $(\downarrow)$ \\
\colrule
Latent prediction (JEPA) & $\bm{0.0020 \pm 0.0001}$ \\
Reconstruction (MAE)     & $0.0027 \pm 0.0000$ \\
Random encoder & $0.0040 \pm 0.0003$ \\
Time-resolved observables & $0.0024 \pm 0.0000$ \\
Time-averaged observables     &  $0.0416 \pm 0.0006$ \\
\end{tabular}
\end{ruledtabular}
\end{table}

\subsection{Imaginary-time relaxation}
\label{sec:imag}

Both settings above are unitary quenches. The evaluation pipeline applies directly to trajectories generated by imaginary-time evolution, $\vert{}\psi(\tau)\rangle \propto e^{-H\tau}\vert{}\psi_0\rangle$ (Section~\ref{sec:hamiltonian}), a non-unitary dynamics that models hardware-implemented quantum imaginary-time algorithms~\cite{motta2020determining,mcardle2019variational}.

The ordering of the two objectives is unchanged. For the field map alone (Table~\ref{tab:imag}), latent prediction achieves an error of $0.0019\pm0.0000$ vs $0.0031\pm0.0001$ for reconstruction (a $1.6$-fold advantage). Hence, its benefit is not restricted to unitary dynamics. This regime also tests our interpretation of the time-averaged control. While it recovers the field only to $0.456$ under real-time quenches (Table~\ref{tab:joint}), it achieves $0.0574\pm0.0005$ here, eight times lower. Because relaxation trajectories terminate at a Hamiltonian-dependent ground state, static observables retain parameter information even when the time dependence is discarded. The time-resolved observables reach only $0.0204\pm0.0014$, eleven times the error of latent prediction. The random encoder reaches $0.0032\pm0.0003$, a factor of $1.7$ above latent prediction and indistinguishable from reconstruction.

On the joint task in imaginary time (last three columns of Table~\ref{tab:joint}), latent prediction reaches an overall error of $0.1547\pm0.0041$ against $0.2453\pm0.0046$ for reconstruction, a factor of $1.6$, and is the most accurate input in every column. On the field, the time-resolved observables, the most accurate input in real time, reach twice the error of latent prediction. The random encoder reaches $0.266\pm0.027$, a factor of $1.7$ above latent prediction and again not distinguishable from reconstruction. The error of the time-averaged control falls on the field, from $0.456$ in real time to $0.193$, but not on the couplings ($0.82$ to $0.92$).

\begin{table}[h]
\caption{\label{tab:imag}
Field-recovery error under imaginary-time evolution, setup otherwise as in Sec.~\ref{sec:results}. Best in bold.}
\begin{ruledtabular}
\begin{tabular}{lc}
Representation & Recovery error $(\downarrow)$ \\
\colrule
Latent prediction (JEPA) & $\bm{0.0019 \pm 0.0000}$ \\
Reconstruction (MAE)     & $0.0031 \pm 0.0001$ \\
Random encoder & $0.0032 \pm 0.0003$ \\
Time-resolved observables & $0.0204 \pm 0.0014$ \\
Time-averaged observables     &  $0.0574 \pm 0.0005$ \\
\end{tabular}
\end{ruledtabular}
\end{table}

\section{Discussion}
\label{sec:discussion}

With the same tuning budget (Appendix~\ref{app:fairness}), latent prediction outperforms reconstruction in all four settings, improving recovery accuracy by factors of $1.3$ to $1.7$. Both learned representations also perform considerably better than simply averaging the measurements over time, with improvements ranging from $4.4$ to $30$ for latent prediction. This is particularly interesting because the time-averaged control uses the same observables with reduced shot noise (Sec.~\ref{sec:probe}). The improvement therefore highlights the importance of temporal information for identifying the underlying Hamiltonian. However, we find that much of this advantage can already be gained without pretraining: a random encoder that preserves the time dependence performs substantially better than the time-averaged control. The additional benefit of learning the representation is nevertheless clear, with latent prediction improving over the random encoder by factors of $1.7$ to $2.1$, compared with $1.0$ to $1.5$ for reconstruction. Interestingly, the time-resolved observables themselves can sometimes be more informative than the learned representations. In the joint real-time task, they allow more accurate recovery of local fields than either encoder (Table~\ref{tab:joint}). This suggests that local fields can be identified effectively from the dynamics of individual spins, whereas recovering interactions between sites benefits more directly from learning their collective dynamics.

The separation between the two objectives is consistent with the argument of Sec.~\ref{sec:objectives}: shot noise is drawn independently for each frame, and retaining it can lower a reconstruction loss but not a cross-window prediction loss. Part of the separation is a matter of scale, not content. The regression head is not scale-invariant, and only latent prediction fixes the scale of its features, through the variance term of Eq.~\eqref{eq:vicreg}. With every feature channel standardized over the training trajectories, reconstruction improves in all four settings, latent prediction changes less, and the factors separating the two fall from between $1.3$ and $1.7$ to between $1.2$ and $1.3$. The ordering survives, and the smaller factors are the better measure of the separation. The regenerated dynamics of Sec.~\ref{sec:dynamics} gives the recovery error a physical scale, and the ordering of the two objectives carries over to them, with a factor of $1.12$ in the occupations against $1.29$ in the root-mean-square standardized error of the parameters.

Since pretraining needs no labels, the encoder can be trained on the raw output of a device and the regression head refitted as the calibration drifts~\cite{proctor2020detecting,kelly2018physical}. The same pipeline applies to non-unitary dynamics and could target spatially resolved maps of dissipation~\cite{bairey2020learning,samach2022lindblad}. Whether a regression head fitted on classically tractable systems transfers to the records of larger hardware is the next test.

\section{Conclusion}
\label{sec:conclusion}

We proposed a latent world model for learning Hamiltonians from measured quantum dynamics. The model is a JEPA trained on one objective, to predict the representation of the next window of measurements from the previous one. Because the Hamiltonian is shared between windows and the shot noise is not, this objective retains the dynamics and discards the noise.

We implemented the model as a spatiotemporal convolutional encoder with a predictor, pretrained it without labels on finite-shot observables with detection error, and read out the Hamiltonian parameters from the frozen encoder with a pointwise regression head. We benchmarked it on a 2D transverse-field Ising lattice in real and imaginary time, simulated exactly, and on a disordered Heisenberg chain of $100$ sites, simulated with tensor networks, against a MAE, the same encoder left untrained, and regression directly on the observables.

We achieved the lowest recovery error of all representations in all four settings, for both the fields and the couplings. The advantage over the MAE is a factor of $1.3$ to $1.7$, and the advantage over regression on time-averaged observables is a factor of $4.4$ to $30$. The recovered Hamiltonian reproduces the measured dynamics. Fitted on the first $8$ frames of a record, it predicts the remaining $24$ to within $1.2$ times their shot noise, and $2.7$ times more accurately than the predictor of the world model rolled forward without a Hamiltonian. Latent prediction is thus a practical self-supervised objective for learning quantum dynamics from measurements.

\begin{acknowledgments}
O.\,K. acknowledges the support from UK EPSRC award under the Agreement No. EP/Z53318X/1 (QCi3 Hub) and the UKRI Robotics funding on AI for Quantum. The authors acknowledge IT Services at The University of Sheffield for the provision of computing services and using Stanage HPC.
\end{acknowledgments}

\vspace{14pt}

\emph{AI statement.} The authors used Claude Fable 5.1 for language editing and code polishing during the preparation of this manuscript. All AI-assisted outputs were verified by the authors, who assume full responsibility for the accuracy and integrity of the work.

\appendix

\section{Inhomogeneous-parameter ensembles}
\label{app:fields}

Each trajectory draws an independent smooth target map, for the Ising model of Eq.~\eqref{eq:tfim}. For the field task we generate $h_i$ as a low-frequency surface,
\begin{equation}
\tilde{h}(x,y) \;=\; a_0 + a_x\,x + a_y\,y + a_{xy}\,xy ,
\label{eq:field}
\end{equation}
with $(x,y)$ the normalized site coordinates on $[-1,1]^2$ and the four coefficients drawn independently and uniformly. The surface is then affinely rescaled to span $[h_{\min},h_{\max}]$. This yields a structured but low-dimensional field, smoothly varying across the lattice rather than site-wise random, so that the recovery target has spatial coherence while remaining nontrivial to infer.

For the joint task each trajectory draws, besides the field surface, two independent coupling surfaces of the form~\eqref{eq:field}, one for the horizontal and one for the vertical bonds, each rescaled to $[J_{\min},J_{\max}]$, so that the field and both bond directions vary at once. On a periodic lattice with $L_x,L_y>2$ the bonds tile a $2\times L_x\times L_y$ grid, so that the coupling map is a coherent two-channel surface.

In the reported experiments the field task uses $[h_{\min},h_{\max}]=[0.5,2.0]$ with fixed $J=0.3$, so that the field exceeds the coupling on every bond. The joint task uses $[h_{\min},h_{\max}]=[0.5,2.0]$ with $[J_{\min},J_{\max}]=[0.3,1.0]$, chosen so that neither parameter dominates the other: the bond correlators retain their sensitivity to $J$ while the single-site precession retains its sensitivity to $h$. The low-dimensional ensemble ensures that the governing degrees of freedom are far fewer than the dimension of the observable trajectory, so that the recovery task tests a representation's ability to compress the dynamics toward its generator rather than to memorize high-dimensional detail.

The Heisenberg chain of Eq.~\eqref{eq:heis} breaks this design: its on-site fields are drawn independently, $h_i\sim\mathcal{U}[-1,1]$, exactly the ensemble of Ref.~\cite{wilde2026scalably}, as in the disordered chains studied in the context of many-body localization~\cite{schreiber2015observation,smith2016many,choi2016exploring,abanin2019colloquium}, so the target has no spatial smoothness to exploit and the number of governing degrees of freedom equals the number of sites. The two ensembles therefore cover two limits: a strongly compressible target on a small lattice, and an incompressible one on a large chain.

\section{Network architectures}
\label{app:arch}

The shared encoder (Section~\ref{sec:objectives}) is a 3D ConvNeXt-style residual network~\cite{liu2022convnet}, a convolutional network in which each block adds its output to its input, operating on frame windows of shape $(F, L_x, L_y)$. Input features comprise a single occupation channel and two bond-correlator channels on the lattice, and the four channels of Section~\ref{sec:measurement} on the chain, with windows extracted at a stride of $4$ frames along each trajectory on the lattice. On the chain, where $T=2F$, there is a single pair of context and target windows. Each block applies a spatiotemporal convolution to every channel separately, normalizes the result~\cite{ba2016layer}, and passes it through a two-layer network that acts on each site independently and widens the channel dimension before narrowing it again. In the random-encoder experiments, the output of every residual branch is given unit weight, instead of the initial weight of $10^{-6}$ used in training, so that every block contributes. All models are optimized using AdamW~\cite{loshchilov2018decoupled} with a weight decay of $0.05$ for $1000$ pretraining epochs in mixed precision and $1000$ epochs for the regression head, using batch sizes of $128$ for the lattice and $64$ for the Heisenberg chain, on which the fit of the head is stopped early, once its validation error has not improved for $100$ epochs. Pretraining gradients are clipped to a global norm of unity, which bounds the step at high learning rates. Encoder checkpoints are evaluated every ten epochs on a $15\%$ held-out split of the pretraining pool using the unsupervised objective loss, requiring no target labels. The encoder uses channel widths of $(32, 64, 128)$ across three stages with two residual blocks per stage. It omits spatial downsampling to preserve full spatial resolution for per-site and per-bond target recovery, avoiding information loss from pooling. Local convolutions match the bounded speed at which correlations spread in short-range lattice systems~\cite{lieb1972finite,cheneau2012light}. Omitting spatial striding also allows the identical network to run on the 1D Heisenberg chain ($L_x=1, L_y=n$) without collapsing spatial axes. The resulting output feature map has shape $(128, F, L_x, L_y)$: the time axis keeps its $F$ frames.

The three VICReg loss terms [Eq.~\eqref{eq:vicreg}] are evaluated across the channel dimension at every spatiotemporal position of the feature map, using weighting coefficients $(\lambda,\mu,\nu)=(2,20,2)$. For the MAE, a space-time tube mask obscures ten of the twelve lattice sites per window, corresponding to a masking ratio of $0.85$ (an interior optimum, Table~\ref{tab:mask}). Because these settings are objective-specific, the sweep behind the mask ratio is given in Appendix~\ref{app:fairness}.

\begin{figure}[t]
\centering
\includegraphics[width=\columnwidth]{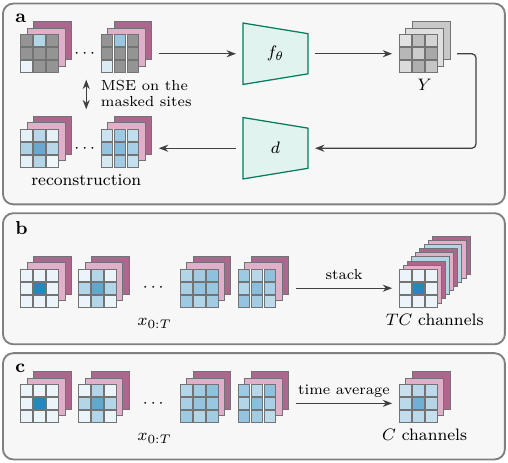}
\caption{\label{fig:pipeline_mae_raw}%
The three comparison representations, using the conventions of Fig.~\ref{fig:pipeline}. (a) Masked autoencoder pretraining: the shared encoder $f_\theta$ receives a window whose sites are hidden by a space-time tube mask (grey), a decoder $d$ maps the representation $Y$ back to observable space, and the loss is the mean-squared error on the masked sites alone. It is thus evaluated within a single window and in observable space, whereas the JEPA objective of Fig.~\ref{fig:pipeline}(b) is evaluated between one window and the next, in latent space. (b) The time-resolved observables, which hand all $T$ frames at a site to the regression head as $TC$ channels, with no encoder. (c) The time-averaged control, which replaces the learned encoder by a time average of the observables ($C$ channels) and performs no representation learning, and which therefore measures how much of the governing map is recoverable from a purely static description of the trajectory. The representation at the right of each panel is passed to the identical regression head of Fig.~\ref{fig:pipeline}(c), so that the rows of every table in this work are directly comparable.}
\end{figure}

The two methods attach different pretraining heads to the shared encoder. For the JEPA, the predictor $g$ is a convolutional network that widens and then narrows the channel dimension and maps latent feature maps into representations of identical shape without returning to observable space. The target representation $z'=f(\text{target})$ is produced by applying the encoder to the future window. In contrast to common joint-embedding setups~\cite{grill2020bootstrap,chen2021exploring}, we let gradients flow through the target branch as well as the context branch, rather than freezing the target branch, as the variance and covariance terms suffice to prevent collapse. This removes a design asymmetry between objectives. For the MAE, the decoder maps encoder outputs back to observable space to reconstruct masked values. Because the encoder performs no spatial downsampling, the decoder requires no upsampling and reconstructs at native lattice resolution. Fig.~\ref{fig:pipeline_mae_raw} depicts the pipeline architectures omitted from the main text.

The shared encoder contains $373{,}504$ trainable parameters, while the auxiliary pretraining heads comprise $2{,}303{,}360$ parameters for the JEPA predictor and $387$ parameters for the default MAE decoder (a single $1\times1$ output convolution requiring no spatial upsampling). The regression head operating on the encoder feature map uses $12{,}481$ parameters for field recovery, whereas the head of the raw control requires $4{,}481$ parameters, with the joint head adding $65$ parameters per additional output channel. For the 1D Heisenberg chain, input channel differences yield parameter counts of $374{,}368$ for the encoder, $516$ for the decoder, and $4{,}545$ for the raw head. The structural asymmetry between discarded pretraining heads is swept in Appendix~\ref{app:fairness}.

\section{The end-to-end procedure}
\label{app:procedure}

The experimental workflow applies uniformly across all Hamiltonians, tasks, and readout protocols, and divides into a simulation stage and a learning stage:

\begin{enumerate}
\item Sample independent parameter maps per trajectory (smooth surfaces for TFIM, i.i.d.\ disorder for Heisenberg chains).

\item Evolve initial states over $T$ uniform time frames using exact statevector propagation (TFIM) or TEBD (chain).

\item Convert states into noisy finite-shot trajectory tensors $(T, L_x, L_y, C)$ and target maps. Lattice readouts use Born-rule bitstring sampling to preserve channel correlations. Chain readouts apply binomial shot noise $k_+ \sim \mathrm{Bin}\big(S, (1+e')/2\big)$ to error-scaled expectations $e' = (1-2\epsilon)^m e$, giving each channel $S$ shots independently. Pre-computed trajectories are cached for bit-identical evaluation across models.

\item Pretrain JEPA and MAE using unlabelled trajectory tensors, both under the same seed. The two encoders are drawn one after the other from a single seeded random stream, so they share the seed but start from different weights.

\item Freeze encoders, train downstream regression heads on labelled sets, and benchmark recovery errors on held-out test trajectories. The raw control baseline is optimized under identical hyperparameter search protocols.
\end{enumerate}

\section{Configuration of the reported runs}
\label{app:config}

All runs reported in this work share the data protocol, optimizer settings, epoch budgets, and label-free validation selection mechanisms of Appendices~\ref{app:arch} and~\ref{app:procedure}, across four random seeds ($\{0, 1, 2, 3\}$). One adjustment limits direct comparison across tables: encoder windows use $F=8$ frames for lattice targets versus $F=16$ for the $100$-site Heisenberg chain, with batch sizes adjusted from $128$ to $64$ to accommodate larger frames, and with the fit of the head on the chain stopped early (Appendix~\ref{app:arch}). Every rate quoted below is otherwise selected in the same way for every target: the full $19$-point pretraining grid and $50$-point head-rate grid of Appendix~\ref{app:fairness}, with the control head rate refitted over that same grid at every seed.

\begin{table*}[t]
\caption{\label{tab:config}
Learning rates and control convention behind every table of recovery errors in this work, quoted as the pair $(\mathrm{lr},\mathrm{lr}_{\mathrm{head}})$ of pretraining rate and head rate. Every other setting is shared and is listed in the text of this appendix. The control has no pretraining stage, so only a head rate is quoted for it. Every rate in this table is selected by the search of Appendix~\ref{app:fairness}. The decoder and mask sweeps of Tables~\ref{tab:decoder} and~\ref{tab:mask} move reconstruction alone and have no latent-prediction entry. N/A: Not applicable.}
\begin{ruledtabular}
\begin{tabular}{llccc}
Table & Task and readout & Latent prediction (JEPA) & Reconstruction (MAE) & Control \\
\colrule
\ref{tab:joint}    & field and couplings, occupation & $(1\times10^{-4},\,4\times10^{-3})$   & $(9\times10^{-4},\,1\times10^{-2})$   & $1.1\times10^{-2}$ \\
\ref{tab:joint}    & field and couplings, occupation, imaginary time & $(1\times10^{-4},\,2\times10^{-3})$   & $(3\times10^{-3},\,9\times10^{-3})$   & $1.3\times10^{-2}$ \\
\ref{tab:imag}     & field, occupation, imaginary time & $(1\times10^{-4},\,4\times10^{-3})$ & $(1\times10^{-3},\,7\times10^{-3})$ & $1.4\times10^{-2}$ \\
\ref{tab:heis}     & chain disorder field, $F=16$    & $(1\times10^{-4},\,3\times10^{-3})$   & $(8\times10^{-4},\,1.2\times10^{-2})$ & $2.3\times10^{-2}$ \\
\ref{tab:decoder}  & field, occupation, decoder sweep & N/A  & $(6\times10^{-3},\,1.5\times10^{-2})$ & N/A \\
\ref{tab:mask}     & field, occupation, mask sweep   & N/A                    & $(6\times10^{-3},\,1.5\times10^{-2})$ & N/A \\
\end{tabular}
\end{ruledtabular}
\end{table*}

The downstream regression head collapses temporal structure twice: once within each window and again across the $T/F$ trajectory windows. Because i.i.d.\ disorder fields encode parameter information primarily in temporal dynamics without spatial coherence, extending the encoder window to $F=16$ halves the outer window count from four to two, preserving higher-frequency temporal features. As expected for a lower-dimensional representation, the selected head learning rate is higher for the control ($1.1\times 10^{-2}$ to $2.3\times 10^{-2}$) than for the learned encoders ($2.0\times 10^{-3}$ to $1.2\times 10^{-2}$) on all four targets. Because the control omits pretraining, its performance does not depend on the pretraining run, and reported values are extracted from latent-prediction runs.

\section{Tuning protocol}
\label{app:fairness}

Everything the representations share, from the architecture and the data to the epoch budget and the label-free checkpoint rule, is listed in Appendices~\ref{app:arch} and~\ref{app:config}. The objective and the two learning rates are not shared, and are selected per representation. This appendix records how that selection is made and which additional settings were swept.

The pretraining learning rate and the head learning rate are chosen in two stages, as the head rate does not enter pretraining. The encoder is pretrained once per pretraining rate at a single seed, and a fresh regression head is then fitted on each saved encoder at every head learning rate. The pretraining grid holds $19$ values, $1\times10^{-4}$ to $9\times10^{-4}$ in steps of $10^{-4}$ and $1\times10^{-3}$ to $1\times10^{-2}$ in steps of $10^{-3}$. For latent prediction, pretraining rates of $2\times10^{-3}$ and above ($4\times10^{-3}$ on the chain) do not train in mixed precision and are never selected. The head-rate grid holds $50$ values, $10^{-3}$ to $5\times10^{-2}$ in steps of $10^{-3}$. Selection uses the validation error of the head alone, and the test error is reported for the selected pair. The control has no pretraining stage, so only its head learning rate is meaningful, and that learning rate is obtained over the same grid at every seed of the latent-prediction run. The selected pairs are collected in Table~\ref{tab:config} and are then rerun over the four seeds. Those runs give the values reported in Section~\ref{sec:results}.

The predictor and decoder heads discarded after pretraining differ in size by nearly four orders of magnitude (Appendix~\ref{app:arch}). Table~\ref{tab:decoder} summarizes the MAE with progressively deeper decoders, up to one that approximately matches the predictor's size, at reconstruction's own tuned field-task learning rates. The error rises monotonically with decoder depth, so reconstruction is already reported at its best decoder on that task.

\begin{table}[t]
\caption{\label{tab:decoder}
MAE recovery error versus decoder capacity, setup as in Section~\ref{sec:results}. $n_{\mathrm{blocks}}$ ConvNeXt refinement blocks precede the output convolution, and $n_{\mathrm{blocks}}=16$ approximately matches the JEPA predictor's parameter count. Row $0$ is the default. All rows run at reconstruction's tuned field-task rates (Appendix~\ref{app:config}), so they differ only in decoder capacity. Four seeds, best in bold.}
\begin{ruledtabular}
\begin{tabular}{lcc}
Decoder & Parameters & MAE recovery error $(\downarrow)$ \\
\colrule
$n_{\mathrm{blocks}}=0$ (default) & $387$           & $\bm{0.0038 \pm 0.0009}$ \\
$n_{\mathrm{blocks}}=2$           & $271{,}747$     & $0.0042 \pm 0.0009$ \\
$n_{\mathrm{blocks}}=4$           & $543{,}107$     & $0.0055 \pm 0.0015$ \\
$n_{\mathrm{blocks}}=16$          & $2{,}171{,}267$ & $0.0150 \pm 0.0040$ \\
\end{tabular}
\end{ruledtabular}
\end{table}

Reconstruction has one setting with no JEPA counterpart, the tube-mask ratio~\cite{he2022masked,feichtenhofer2022masked}, just as latent prediction has two settings with no reconstruction counterpart, the coefficients of Eq.~\eqref{eq:vicreg} and the geometry of the predictor. Table~\ref{tab:mask} sweeps the mask ratio at the same tuned rates. The error falls to a minimum at $10$ of $12$ sites and rises steeply at eleven. Latent prediction does not use the mask and is identical in every experiment.

\begin{table}[t]
\caption{\label{tab:mask}
MAE recovery error versus tube-mask ratio, the fraction of lattice sites hidden, setup as in Section~\ref{sec:results}. The sweep runs at reconstruction's tuned field-task settings (Appendix~\ref{app:config}) over four seeds. With $n_{\mathrm{mask}}=\mathrm{round}(r\,n_{\mathrm{sites}})$ and $n_{\mathrm{sites}}=12$, the ratio resolves in steps of $1/12$, one swept value per site count. Best in bold.}
\begin{ruledtabular}
\begin{tabular}{lcc}
Mask ratio & Sites hidden & MAE recovery error $(\downarrow)$ \\
\colrule
$0.50$           & $6/12$  & $0.0114 \pm 0.0043$ \\
$0.60$           & $7/12$  & $0.0067 \pm 0.0010$ \\
$0.70$           & $8/12$  & $0.0067 \pm 0.0037$ \\
$0.75$           & $9/12$  & $0.0044 \pm 0.0007$ \\
$0.85$ (default) & $10/12$ & $\bm{0.0038 \pm 0.0009}$ \\
$0.90$           & $11/12$ & $0.0279 \pm 0.0096$ \\
\end{tabular}
\end{ruledtabular}
\end{table}

\bibliography{references}

@article{bernien2017probing,
  title={Probing many-body dynamics on a 51-atom quantum simulator},
  author={Bernien, Hannes and Schwartz, Sylvain and Keesling, Alexander and Levine, Harry and Omran, Ahmed and Pichler, Hannes and Choi, Soonwon and Zibrov, Alexander S and Endres, Manuel and Greiner, Markus and others},
  journal={Nature},
  volume={551},
  number={7682},
  pages={579--584},
  year={2017},
  publisher={Nature Publishing Group UK London},
  doi={10.1038/nature24622},
  url={https://doi.org/10.1038/nature24622}
}

@article{zhang2017observation,
  title={Observation of a many-body dynamical phase transition with a 53-qubit quantum simulator},
  author={Zhang, Jiehang and Pagano, Guido and Hess, Paul W and Kyprianidis, Antonis and Becker, Patrick and Kaplan, Harvey and Gorshkov, Alexey V and Gong, Zhe-Xuan and Monroe, Christopher},
  journal={Nature},
  volume={551},
  number={7682},
  pages={601--604},
  year={2017},
  publisher={Nature Publishing Group UK London},
  doi={10.1038/nature24654},
  url={https://doi.org/10.1038/nature24654}
}

@article{ebadi2021quantum,
  title={Quantum phases of matter on a 256-atom programmable quantum simulator},
  author={Ebadi, Sepehr and Wang, Tout T and Levine, Harry and Keesling, Alexander and Semeghini, Giulia and Omran, Ahmed and Bluvstein, Dolev and Samajdar, Rhine and Pichler, Hannes and Ho, Wen Wei and others},
  journal={Nature},
  volume={595},
  number={7866},
  pages={227--232},
  year={2021},
  publisher={Nature Publishing Group UK London},
  doi={10.1038/s41586-021-03582-4},
  url={https://doi.org/10.1038/s41586-021-03582-4}
}

@article{scholl2021quantum,
  title={Quantum simulation of 2D antiferromagnets with hundreds of Rydberg atoms},
  author={Scholl, Pascal and Schuler, Michael and Williams, Hannah J and Eberharter, Alexander A and Barredo, Daniel and Schymik, Kai-Niklas and Lienhard, Vincent and Henry, Louis-Paul and Lang, Thomas C and Lahaye, Thierry and others},
  journal={Nature},
  volume={595},
  number={7866},
  pages={233--238},
  year={2021},
  publisher={Nature Publishing Group UK London},
  doi={10.1038/s41586-021-03585-1},
  url={https://doi.org/10.1038/s41586-021-03585-1}
}

@misc{Mendessantos2026,
      title={Exploring the Relaxation Landscape of a 2D Quantum Magnet on a 256-Qubit Processor}, 
      author={Tiago Mendes-Santos and Joseph Vovrosh and Sergi Julià-Farré and Dorian Claveau and Guillaume Villaret and Lucas Béguin and Lucas Leclerc and Laurin Brunner and Wladislaw Krinitsin and Matthias Hecker and Fergus Hayes and Boris Albrecht and Lilian Bourachot and Clémence Briosne-Frejaville and Antoine Cornillot and Julius de Hond and Djibril Diallo and Clément Dupays and Robin Dupont and Thomas Eritzpokhoff and Loïc Henriet and Lucas Lassablière and Arvid Lindberg and Yohann Machu and Hadriel Mamann and Thomas Pansiot and Julien Ripoll and Bruno Ximenez and Henrique Silvério and Joseph Tindall and Markus Schmitt and Markus Heyl and Adrien Signoles and Constantin Dalyac and Antoine Browaeys and Alexandre Dauphin},
      year={2026},
      eprint={2608.07178},
      archivePrefix={arXiv},
      primaryClass={quant-ph},
      url={https://arxiv.org/abs/2608.07178}, 
}

@article{huang2020predicting,
  title={Predicting many properties of a quantum system from very few measurements},
  author={Huang, Hsin-Yuan and Kueng, Richard and Preskill, John},
  journal={Nature Physics},
  volume={16},
  number={10},
  pages={1050--1057},
  year={2020},
  doi={10.1038/s41567-020-0932-7}
}

@article{elben2023randomized,
  title={The randomized measurement toolbox},
  author={Elben, Andreas and Flammia, Steven T. and Huang, Hsin-Yuan and Kueng, Richard and Preskill, John and Vermersch, Beno{\^\i}t and Zoller, Peter},
  journal={Nature Reviews Physics},
  volume={5},
  number={1},
  pages={9--24},
  year={2023},
  doi={10.1038/s42254-022-00535-2}
}

@misc{bermejo2026tnbp,
      title={Tensor Networks with Belief Propagation Cannot Feasibly Simulate Google's Quantum Echoes Experiment}, 
      author={Pablo Bermejo and Benjamin Villalonga and Brayden Ware and Guifre Vidal and Aaron Szasz},
      year={2026},
      eprint={2604.15427},
      archivePrefix={arXiv},
      primaryClass={quant-ph},
      url={https://arxiv.org/abs/2604.15427}, 
}

@article{browaeys2020many,
  title={Many-body physics with individually controlled Rydberg atoms},
  author={Browaeys, Antoine and Lahaye, Thierry},
  journal={Nature Physics},
  volume={16},
  number={2},
  pages={132--142},
  year={2020},
  publisher={Nature Publishing Group UK London},
  doi={10.1038/s41567-019-0733-z},
  url={https://doi.org/10.1038/s41567-019-0733-z}
}

@article{gross2017quantum,
  title={Quantum simulations with ultracold atoms in optical lattices},
  author={Gross, Christian and Bloch, Immanuel},
  journal={Science},
  volume={357},
  number={6355},
  pages={995--1001},
  year={2017},
  publisher={American Association for the Advancement of Science},
  doi={10.1126/science.aal3837},
  url={https://doi.org/10.1126/science.aal3837}
}

@article{altman2021quantum,
  title={Quantum simulators: Architectures and opportunities},
  author={Altman, Ehud and Brown, Kenneth R and Carleo, Giuseppe and Carr, Lincoln D and Demler, Eugene and Chin, Cheng and DeMarco, Brian and Economou, Sophia E and Eriksson, Mark A and Fu, Kai-Mei C and others},
  journal={PRX Quantum},
  volume={2},
  number={1},
  pages={017003},
  year={2021},
  publisher={APS},
  doi={10.1103/PRXQuantum.2.017003},
  url={https://doi.org/10.1103/PRXQuantum.2.017003}
}

@article{daley2022practical,
  title={Practical quantum advantage in quantum simulation},
  author={Daley, Andrew J and Bloch, Immanuel and Kokail, Christian and Flannigan, Stuart and Pearson, Natalie and Troyer, Matthias and Zoller, Peter},
  journal={Nature},
  volume={607},
  number={7920},
  pages={667--676},
  year={2022},
  publisher={Nature Publishing Group UK London},
  doi={10.1038/s41586-022-04940-6},
  url={https://doi.org/10.1038/s41586-022-04940-6}
}

@article{arute2019quantum,
  title={Quantum supremacy using a programmable superconducting processor},
  author={Arute, Frank and Arya, Kunal and Babbush, Ryan and Bacon, Dave and Bardin, Joseph C and Barends, Rami and Biswas, Rupak and Boixo, Sergio and Brand{\~a}o, Fernando GSL and Buell, David A and others},
  journal={Nature},
  volume={574},
  number={7779},
  pages={505--510},
  year={2019},
  publisher={Nature Publishing Group UK London},
  doi={10.1038/s41586-019-1666-5},
  url={https://doi.org/10.1038/s41586-019-1666-5}
}

@article{kim2023evidence,
  title={Evidence for the utility of quantum computing before fault tolerance},
  author={Kim, Youngseok and Eddins, Andrew and Anand, Sajant and Wei, Ken Xuan and Van Den Berg, Ewout and Rosenblatt, Sami and Nayfeh, Hasan and Wu, Yantao and Zaletel, Michael and Temme, Kristan and others},
  journal={Nature},
  volume={618},
  number={7965},
  pages={500--505},
  year={2023},
  publisher={Nature Publishing Group UK London},
  doi={10.1038/s41586-023-06096-3},
  url={https://doi.org/10.1038/s41586-023-06096-3}
}

@article{bakr2009quantum,
  title={A quantum gas microscope for detecting single atoms in a Hubbard-regime optical lattice},
  author={Bakr, Waseem S and Gillen, Jonathon I and Peng, Amy and F{\"o}lling, Simon and Greiner, Markus},
  journal={Nature},
  volume={462},
  number={7269},
  pages={74--77},
  year={2009},
  publisher={Nature Publishing Group UK London},
  doi={10.1038/nature08482},
  url={https://doi.org/10.1038/nature08482}
}

@article{sherson2010single,
  title={Single-atom-resolved fluorescence imaging of an atomic Mott insulator},
  author={Sherson, Jacob F and Weitenberg, Christof and Endres, Manuel and Cheneau, Marc and Bloch, Immanuel and Kuhr, Stefan},
  journal={Nature},
  volume={467},
  number={7311},
  pages={68--72},
  year={2010},
  publisher={Nature Publishing Group UK London},
  doi={10.1038/nature09378},
  url={https://doi.org/10.1038/nature09378}
}

@article{gross2021quantum,
  title={Quantum gas microscopy for single atom and spin detection},
  author={Gross, Christian and Bakr, Waseem S},
  journal={Nature Physics},
  volume={17},
  number={12},
  pages={1316--1323},
  year={2021},
  publisher={Nature Publishing Group UK London},
  doi={10.1038/s41567-021-01370-5},
  url={https://doi.org/10.1038/s41567-021-01370-5}
}

@article{schreiber2015observation,
  title={Observation of many-body localization of interacting fermions in a quasirandom optical lattice},
  author={Schreiber, Michael and Hodgman, Sean S and Bordia, Pranjal and L{\"u}schen, Henrik P and Fischer, Mark H and Vosk, Ronen and Altman, Ehud and Schneider, Ulrich and Bloch, Immanuel},
  journal={Science},
  volume={349},
  number={6250},
  pages={842--845},
  year={2015},
  publisher={American Association for the Advancement of Science},
  doi={10.1126/science.aaa7432},
  url={https://doi.org/10.1126/science.aaa7432}
}

@article{smith2016many,
  title={Many-body localization in a quantum simulator with programmable random disorder},
  author={Smith, Jacob and Lee, Aaron and Richerme, Philip and Neyenhuis, Brian and Hess, Paul W and Hauke, Philipp and Heyl, Markus and Huse, David A and Monroe, Christopher},
  journal={Nature Physics},
  volume={12},
  number={10},
  pages={907--911},
  year={2016},
  publisher={Nature Publishing Group UK London},
  doi={10.1038/nphys3783},
  url={https://doi.org/10.1038/nphys3783}
}

@article{choi2016exploring,
  title={Exploring the many-body localization transition in two dimensions},
  author={Choi, Jae-yoon and Hild, Sebastian and Zeiher, Johannes and Schau{\ss}, Peter and Rubio-Abadal, Antonio and Yefsah, Tarik and Khemani, Vedika and Huse, David A and Bloch, Immanuel and Gross, Christian},
  journal={Science},
  volume={352},
  number={6293},
  pages={1547--1552},
  year={2016},
  publisher={American Association for the Advancement of Science},
  doi={10.1126/science.aaf8834},
  url={https://doi.org/10.1126/science.aaf8834}
}

@article{abanin2019colloquium,
  title={Colloquium: Many-body localization, thermalization, and entanglement},
  author={Abanin, Dmitry A and Altman, Ehud and Bloch, Immanuel and Serbyn, Maksym},
  journal={Reviews of Modern Physics},
  volume={91},
  number={2},
  pages={021001},
  year={2019},
  publisher={APS},
  doi={10.1103/RevModPhys.91.021001},
  url={https://doi.org/10.1103/RevModPhys.91.021001}
}

@inproceedings{haah2016sample,
  title={Sample-optimal tomography of quantum states},
  author={Haah, Jeongwan and Harrow, Aram W and Ji, Zhengfeng and Wu, Xiaodi and Yu, Nengkun},
  booktitle = {Proceedings of the 48th Annual ACM Symposium on Theory of Computing (STOC)},
  publisher = {ACM},
  pages={913--925},
  year={2016},
  doi={10.1145/2897518.2897585},
  url={https://doi.org/10.1145/2897518.2897585}
}

@article{cramer2010efficient,
  title={Efficient quantum state tomography},
  author={Cramer, Marcus and Plenio, Martin B and Flammia, Steven T and Somma, Rolando and Gross, David and Bartlett, Stephen D and Landon-Cardinal, Olivier and Poulin, David and Liu, Yi-Kai},
  journal={Nature Communications},
  volume={1},
  number={1},
  pages={149},
  year={2010},
  publisher={Nature Publishing Group UK London},
  doi={10.1038/ncomms1147},
  url={https://doi.org/10.1038/ncomms1147}
}

@article{torlai2018neural,
  title={Neural-network quantum state tomography},
  author={Torlai, Giacomo and Mazzola, Guglielmo and Carrasquilla, Juan and Troyer, Matthias and Melko, Roger and Carleo, Giuseppe},
  journal={Nature physics},
  volume={14},
  number={5},
  pages={447--450},
  year={2018},
  publisher={Nature Publishing Group UK London},
  doi={10.1038/s41567-018-0048-5},
  url={https://doi.org/10.1038/s41567-018-0048-5}
}

@article{da2011practical,
  title={Practical characterization of quantum devices without tomography},
  author={da Silva, Marcus P and Landon-Cardinal, Olivier and Poulin, David},
  journal={Physical Review Letters},
  volume={107},
  number={21},
  pages={210404},
  year={2011},
  publisher={APS},
  doi={10.1103/PhysRevLett.107.210404},
  url={https://doi.org/10.1103/PhysRevLett.107.210404}
}

@article{granade2012robust,
  title={Robust online Hamiltonian learning},
  author={Granade, Christopher E and Ferrie, Christopher and Wiebe, Nathan and Cory, David G},
  journal={New Journal of Physics},
  volume={14},
  number={10},
  pages={103013},
  year={2012},
  publisher={IOP Publishing},
  doi={10.1088/1367-2630/14/10/103013},
  url={https://doi.org/10.1088/1367-2630/14/10/103013}
}

@article{wiebe2014hamiltonian,
  title={Hamiltonian learning and certification using quantum resources},
  author={Wiebe, Nathan and Granade, Christopher and Ferrie, Christopher and Cory, David G},
  journal={Physical Review letters},
  volume={112},
  number={19},
  pages={190501},
  year={2014},
  publisher={APS},
  doi={10.1103/PhysRevLett.112.190501},
  url={https://doi.org/10.1103/PhysRevLett.112.190501}
}

@article{bairey2019learning,
  title={Learning a local Hamiltonian from local measurements},
  author={Bairey, Eyal and Arad, Itai and Lindner, Netanel H},
  journal={Physical Review letters},
  volume={122},
  number={2},
  pages={020504},
  year={2019},
  publisher={APS},
  doi={10.1103/PhysRevLett.122.020504},
  url={https://doi.org/10.1103/PhysRevLett.122.020504}
}

@article{qi2019determining,
  title={Determining a local Hamiltonian from a single eigenstate},
  author={Qi, Xiao-Liang and Ranard, Daniel},
  journal={Quantum},
  volume={3},
  pages={159},
  year={2019},
  publisher={Verein zur F{\"o}rderung des Open Access Publizierens in den Quantenwissenschaften},
  doi={10.22331/q-2019-07-08-159},
  url={https://doi.org/10.22331/q-2019-07-08-159}
}

@article{anshu2021sample,
  title={Sample-efficient learning of interacting quantum systems},
  author={Anshu, Anurag and Arunachalam, Srinivasan and Kuwahara, Tomotaka and Soleimanifar, Mehdi},
  journal={Nature Physics},
  volume={17},
  number={8},
  pages={931--935},
  year={2021},
  publisher={Nature Publishing Group UK London},
  doi={10.1038/s41567-021-01232-0},
  url={https://doi.org/10.1038/s41567-021-01232-0}
}

@article{huang2023learning,
  title={Learning many-body Hamiltonians with Heisenberg-limited scaling},
  author={Huang, Hsin-Yuan and Tong, Yu and Fang, Di and Su, Yuan},
  journal={Physical Review Letters},
  volume={130},
  number={20},
  pages={200403},
  year={2023},
  publisher={APS},
  doi={10.1103/PhysRevLett.130.200403},
  url={https://doi.org/10.1103/PhysRevLett.130.200403}
}

@article{valenti2019hamiltonian,
  title={Hamiltonian learning for quantum error correction},
  author={Valenti, Agnes and van Nieuwenburg, Evert and Huber, Sebastian and Greplova, Eliska},
  journal={Physical Review Research},
  volume={1},
  number={3},
  pages={033092},
  year={2019},
  publisher={APS},
  doi={10.1103/PhysRevResearch.1.033092},
  url={https://doi.org/10.1103/PhysRevResearch.1.033092}
}

@article{gebhart2023learning,
  title={Learning quantum systems},
  author={Gebhart, Valentin and Santagati, Raffaele and Gentile, Antonio Andrea and Gauger, Erik M and Craig, David and Ares, Natalia and Banchi, Leonardo and Marquardt, Florian and Pezz{\`e}, Luca and Bonato, Cristian},
  journal={Nature Reviews Physics},
  volume={5},
  number={3},
  pages={141--156},
  year={2023},
  publisher={Nature Publishing Group UK London},
  doi={10.1038/s42254-022-00552-1},
  url={https://doi.org/10.1038/s42254-022-00552-1}
}

@article{zubida2021optimal,
  title={Optimal short-time measurements for Hamiltonian learning},
  author={Zubida, Assaf and Yitzhaki, Elad and Lindner, Netanel H and Bairey, Eyal},
  journal={arXiv preprint},
  year={2021},
  doi={10.48550/arXiv.2108.08824},
  url={https://doi.org/10.48550/arXiv.2108.08824}
}

@article{hangleiter2024robustly,
  title={Robustly learning the Hamiltonian dynamics of a superconducting quantum processor},
  author={Hangleiter, Dominik and Roth, Ingo and Fuksa, Jon{\'a}{\v{s}} and Eisert, Jens and Roushan, Pedram},
  journal={Nature Communications},
  volume={15},
  number={1},
  pages={9595},
  year={2024},
  publisher={Nature Publishing Group UK London},
  doi={10.1038/s41467-024-52629-3},
  url={https://doi.org/10.1038/s41467-024-52629-3}
}

@article{stilck2024efficient,
  title={Efficient and robust estimation of many-qubit Hamiltonians},
  author={Stilck Fran{\c{c}}a, Daniel and Markovich, Liubov A and Dobrovitski, Viatcheslav V and Werner, Albert H and Borregaard, Johannes},
  journal={Nature Communications},
  volume={15},
  number={1},
  pages={311},
  year={2024},
  publisher={Nature Publishing Group UK London},
  doi={10.1038/s41467-023-44012-5},
  url={https://doi.org/10.1038/s41467-023-44012-5}
}

@article{carleo2019machine,
  title={Machine learning and the physical sciences},
  author={Carleo, Giuseppe and Cirac, Ignacio and Cranmer, Kyle and Daudet, Laurent and Schuld, Maria and Tishby, Naftali and Vogt-Maranto, Leslie and Zdeborov{\'a}, Lenka},
  journal={Reviews of Modern Physics},
  volume={91},
  number={4},
  pages={045002},
  year={2019},
  publisher={APS},
  doi={10.1103/RevModPhys.91.045002},
  url={https://doi.org/10.1103/RevModPhys.91.045002}
}

@article{carrasquilla2017machine,
  title={Machine learning phases of matter},
  author={Carrasquilla, Juan and Melko, Roger G},
  journal={Nature physics},
  volume={13},
  number={5},
  pages={431--434},
  year={2017},
  publisher={Nature Publishing Group UK London},
  doi={10.1038/nphys4035},
  url={https://doi.org/10.1038/nphys4035}
}

@article{van2017learning,
  title={Learning phase transitions by confusion},
  author={Van Nieuwenburg, Evert PL and Liu, Ye-Hua and Huber, Sebastian D},
  journal={Nature Physics},
  volume={13},
  number={5},
  pages={435--439},
  year={2017},
  publisher={Nature Publishing Group UK London},
  doi={10.1038/nphys4037},
  url={https://doi.org/10.1038/nphys4037}
}

@article{carleo2017solving,
  title={Solving the quantum many-body problem with artificial neural networks},
  author={Carleo, Giuseppe and Troyer, Matthias},
  journal={Science},
  volume={355},
  number={6325},
  pages={602--606},
  year={2017},
  publisher={American Association for the Advancement of Science},
  doi={10.1126/science.aag2302},
  url={https://doi.org/10.1126/science.aag2302}
}

@article{rem2019identifying,
  title={Identifying quantum phase transitions using artificial neural networks on experimental data},
  author={Rem, Benno S and K{\"a}ming, Niklas and Tarnowski, Matthias and Asteria, Luca and Fl{\"a}schner, Nick and Becker, Christoph and Sengstock, Klaus and Weitenberg, Christof},
  journal={Nature Physics},
  volume={15},
  number={9},
  pages={917--920},
  year={2019},
  publisher={Nature Publishing Group UK London},
  doi={10.1038/s41567-019-0554-0},
  url={https://doi.org/10.1038/s41567-019-0554-0}
}

@article{bohrdt2019classifying,
  title={Classifying snapshots of the doped Hubbard model with machine learning},
  author={Bohrdt, Annabelle and Chiu, Christie S and Ji, Geoffrey and Xu, Muqing and Greif, Daniel and Greiner, Markus and Demler, Eugene and Grusdt, Fabian and Knap, Michael},
  journal={Nature Physics},
  volume={15},
  number={9},
  pages={921--924},
  year={2019},
  publisher={Nature Publishing Group UK London},
  doi={10.1038/s41567-019-0565-x},
  url={https://doi.org/10.1038/s41567-019-0565-x}
}

@article{dawid2022modern,
  title={Modern applications of machine learning in quantum sciences},
  author={Dawid, Anna and Arnold, Julian and Requena, Borja and Gresch, Alexander and P{\l}odzie{\'n}, Marcin and Donatella, Kaelan and Nicoli, Kim A and Stornati, Paolo and Koch, Rouven and B{\"u}ttner, Miriam and others},
  journal={arXiv preprint},
  year={2022},
  doi={10.48550/arXiv.2204.04198},
  url={https://doi.org/10.48550/arXiv.2204.04198}
}

@inproceedings{chen2020simple,
  title={A simple framework for contrastive learning of visual representations},
  author={Chen, Ting and Kornblith, Simon and Norouzi, Mohammad and Hinton, Geoffrey},
  booktitle={Proceedings of the 37th International Conference on Machine Learning (ICML)},
  series={Proceedings of Machine Learning Research},
  volume={119},
  pages={1597--1607},
  year={2020},
  publisher={PMLR},
  url={https://proceedings.mlr.press/v119/chen20j.html}
}

@inproceedings{grill2020bootstrap,
  title={Bootstrap your own latent a new approach to self-supervised learning},
  author={Grill, Jean-Bastien and Strub, Florian and Altch{\'e}, Florent and Tallec, Corentin and Richemond, Pierre and Buchatskaya, Elena and Doersch, Carl and Avila Pires, Bernardo and Guo, Zhaohan and Gheshlaghi Azar, Mohammad and others},
  booktitle={Advances in Neural Information Processing Systems (NeurIPS)},
  publisher={Curran Associates, Inc.},
  volume={33},
  pages={21271--21284},
  year={2020},
  url={https://dl.acm.org/doi/10.5555/3495724.3497510}
}

@inproceedings{chen2021exploring,
  title={Exploring simple siamese representation learning},
  author={Chen, Xinlei and He, Kaiming},
  booktitle={Proceedings of the IEEE/CVF Conference on Computer Vision and Pattern Recognition (CVPR)},
  publisher={IEEE},
  pages={15745--15753},
  year={2021},
  doi={10.1109/CVPR46437.2021.01549},
  url={https://doi.org/10.1109/CVPR46437.2021.01549}
}

@inproceedings{zbontar2021barlow,
  title={{B}arlow {T}wins: Self-supervised learning via redundancy reduction},
  author={Zbontar, Jure and Jing, Li and Misra, Ishan and LeCun, Yann and Deny, St{\'e}phane},
  booktitle={Proceedings of the 38th International Conference on Machine Learning (ICML)},
  series={Proceedings of Machine Learning Research},
  volume={139},
  pages={12310--12320},
  year={2021},
  publisher={PMLR},
  url={https://proceedings.mlr.press/v139/zbontar21a.html}
}

@inproceedings{bardes2021vicreg,
  title={{VICReg}: Variance-invariance-covariance regularization for self-supervised learning},
  author={Bardes, Adrien and Ponce, Jean and LeCun, Yann},
  booktitle={International Conference on Learning Representations (ICLR)},
  year={2022},
  url={https://openreview.net/forum?id=xm6YD62D1Ub}
}

@misc{lecun2022path,
  title={A path towards autonomous machine intelligence},
  author={LeCun, Yann},
  howpublished={OpenReview},
  note={Version 0.9.2, 27 June 2022},
  year={2022},
  url={https://openreview.net/pdf?id=BZ5a1r-kVsf}
}

@inproceedings{assran2023self,
  title={Self-supervised learning from images with a joint-embedding predictive architecture},
  author={Assran, Mahmoud and Duval, Quentin and Misra, Ishan and Bojanowski, Piotr and Vincent, Pascal and Rabbat, Michael and LeCun, Yann and Ballas, Nicolas},
  booktitle={Proceedings of the IEEE/CVF Conference on Computer Vision and Pattern Recognition (CVPR)},
  publisher={IEEE},
  pages={15619--15629},
  year={2023},
  doi={10.1109/CVPR52729.2023.01499},
  url={https://doi.org/10.1109/CVPR52729.2023.01499}
}

@article{
bardes2024revisiting,
title={Revisiting Feature Prediction for Learning Visual Representations from Video},
author={Bardes, Adrien and Garrido, Quentin and Ponce, Jean and Chen, Xinlei and Rabbat, Michael and LeCun, Yann and Assran, Mahmoud and Ballas, Nicolas},
journal={Transactions on Machine Learning Research},
issn={2835-8856},
year={2024},
url={https://openreview.net/forum?id=QaCCuDfBk2}
}

@inproceedings{he2022masked,
  title={Masked autoencoders are scalable vision learners},
  author={He, Kaiming and Chen, Xinlei and Xie, Saining and Li, Yanghao and Doll{\'a}r, Piotr and Girshick, Ross},
  booktitle={Proceedings of the IEEE/CVF Conference on Computer Vision and Pattern Recognition (CVPR)}, 
  publisher={IEEE},
  pages={15979--15988},
  year={2022},
  doi={10.1109/CVPR52688.2022.01553},
  url={https://doi.org/10.1109/CVPR52688.2022.01553}
}

@inproceedings{feichtenhofer2022masked,
  title={Masked autoencoders as spatiotemporal learners},
  author={Feichtenhofer, Christoph and Fan, Haoqi and Li, Yanghao and He, Kaiming},
  booktitle={Advances in Neural Information Processing Systems (NeurIPS)},
  publisher={Curran Associates, Inc.},
  volume={35},
  pages={35946--35958},
  year={2022},
  url={https://dl.acm.org/doi/10.5555/3600270.3602875}
}

@inproceedings{tong2022videomae,
  title={Videomae: Masked autoencoders are data-efficient learners for self-supervised video pre-training},
  author={Tong, Zhan and Song, Yibing and Wang, Jue and Wang, Limin},
  booktitle={Advances in Neural Information Processing Systems (NeurIPS)},
  publisher={Curran Associates, Inc.},
  volume={35},
  pages={10078--10093},
  year={2022},
  url={https://dl.acm.org/doi/10.5555/3600270.3601002}
}

@article{balestriero2023cookbook,
  title={A cookbook of self-supervised learning},
  author={Balestriero, Randall and Ibrahim, Mark and Sobal, Vlad and Morcos, Ari and Shekhar, Shashank and Goldstein, Tom and Bordes, Florian and Bardes, Adrien and Mialon, Gregoire and Tian, Yuandong and others},
  journal={arXiv preprint},
  year={2023},
  doi={10.48550/arXiv.2304.12210},
  url={https://doi.org/10.48550/arXiv.2304.12210}
}

@inproceedings{mialon2023self,
  title={Self-supervised learning with lie symmetries for partial differential equations},
  author={Mialon, Gr{\'e}goire and Garrido, Quentin and Lawrence, Hannah and Rehman, Danyal and LeCun, Yann and Kiani, Bobak},
  booktitle={Advances in Neural Information Processing Systems (NeurIPS)},
  publisher={Curran Associates, Inc.},
  volume={36},
  pages={28973--29004},
  year={2023},
  doi={10.52202/075280-1262},
  url={https://doi.org/10.52202/075280-1262}
}

@inproceedings{mccabe2024multiple,
  title={Multiple physics pretraining for spatiotemporal surrogate models},
  author={McCabe, Michael and R{\'e}galdo-Saint Blancard, Bruno and Parker, Liam and Ohana, Ruben and Cranmer, Miles and Bietti, Alberto and Eickenberg, Michael and Golkar, Siavash and Krawezik, Geraud and Lanusse, Francois and others},
  booktitle={Advances in Neural Information Processing Systems (NeurIPS)},
  publisher={Curran Associates, Inc.},
  volume={37},
  pages={119301--119335},
  year={2024},
  url={https://dl.acm.org/doi/10.5555/3737916.3741707}
}

@inproceedings{subramanian2023towards,
  title={Towards foundation models for scientific machine learning: Characterizing scaling and transfer behavior},
  author={Subramanian, Shashank and Harrington, Peter and Keutzer, Kurt and Bhimji, Wahid and Morozov, Dmitriy and Mahoney, Michael W and Gholami, Amir},
  booktitle={Advances in Neural Information Processing Systems (NeurIPS)},
  publisher={Curran Associates, Inc.},
  volume={36},
  pages={71242--71262},
  year={2023},
  doi={10.52202/075280-3119},
  url={https://doi.org/10.52202/075280-3119}
}

@article{brunton2016discovering,
  title={Discovering governing equations from data by sparse identification of nonlinear dynamical systems},
  author={Brunton, Steven L and Proctor, Joshua L and Kutz, J Nathan},
  journal={Proceedings of the national academy of sciences},
  volume={113},
  number={15},
  pages={3932--3937},
  year={2016},
  publisher={National Academy of Sciences},
  doi={10.1073/pnas.1517384113},
  url={https://doi.org/10.1073/pnas.1517384113}
}

@article{raissi2019physics,
  title={Physics-informed neural networks: A deep learning framework for solving forward and inverse problems involving nonlinear partial differential equations},
  author={Raissi, Maziar and Perdikaris, Paris and Karniadakis, George E},
  journal={Journal of Computational physics},
  volume={378},
  pages={686--707},
  year={2019},
  publisher={Elsevier},
  doi={10.1016/j.jcp.2018.10.045},
  url={https://doi.org/10.1016/j.jcp.2018.10.045}
}

@inproceedings{liu2022convnet,
  title={A convnet for the 2020s},
  author={Liu, Zhuang and Mao, Hanzi and Wu, Chao-Yuan and Feichtenhofer, Christoph and Darrell, Trevor and Xie, Saining},
  booktitle={Proceedings of the IEEE/CVF Conference on Computer Vision and Pattern Recognition (CVPR)},
  publisher={IEEE},
  pages={11966--11976},
  year={2022},
  doi={10.1109/CVPR52688.2022.01167},
  url={https://doi.org/10.1109/CVPR52688.2022.01167}
}

@inproceedings{
  loshchilov2018decoupled,
  title={Decoupled Weight Decay Regularization},
  author={Loshchilov, Ilya and Hutter, Frank},
  booktitle={International Conference on Learning Representations (ICLR)},
  year={2019},
  url={https://openreview.net/forum?id=Bkg6RiCqY7},
}

@article{ba2016layer,
  title={Layer normalization},
  author={Ba, Jimmy Lei and Kiros, Jamie Ryan and Hinton, Geoffrey E},
  journal={arXiv preprint},
  year={2016},
  doi={10.48550/arXiv.1607.06450},
  url={https://doi.org/10.48550/arXiv.1607.06450}
}

@article{white1992density,
  title={Density matrix formulation for quantum renormalization groups},
  author={White, Steven R},
  journal={Physical Review letters},
  volume={69},
  number={19},
  pages={2863},
  year={1992},
  publisher={APS},
  doi={10.1103/PhysRevLett.69.2863},
  url={https://doi.org/10.1103/PhysRevLett.69.2863}
}

@article{vidal2003efficient,
  title={Efficient classical simulation of slightly entangled quantum computations},
  author={Vidal, Guifr{\'e}},
  journal={Physical Review letters},
  volume={91},
  number={14},
  pages={147902},
  year={2003},
  publisher={APS},
  doi={10.1103/PhysRevLett.91.147902},
  url={https://doi.org/10.1103/PhysRevLett.91.147902}
}

@article{vidal2004efficient,
  title={Efficient simulation of one-dimensional quantum many-body systems},
  author={Vidal, Guifr{\'e}},
  journal={Physical Review letters},
  volume={93},
  number={4},
  pages={040502},
  year={2004},
  publisher={APS},
  doi={10.1103/PhysRevLett.93.040502},
  url={https://doi.org/10.1103/PhysRevLett.93.040502}
}

@article{schollwock2011density,
  title={The density-matrix renormalization group in the age of matrix product states},
  author={Schollw{\"o}ck, Ulrich},
  journal={Annals of physics},
  volume={326},
  number={1},
  pages={96--192},
  year={2011},
  publisher={Elsevier},
  doi={10.1016/j.aop.2010.09.012},
  url={https://doi.org/10.1016/j.aop.2010.09.012}
}

@article{paeckel2019time,
  title={Time-evolution methods for matrix-product states},
  author={Paeckel, Sebastian and K{\"o}hler, Thomas and Swoboda, Andreas and Manmana, Salvatore R and Schollw{\"o}ck, Ulrich and Hubig, Claudius},
  journal={Annals of Physics},
  volume={411},
  pages={167998},
  year={2019},
  publisher={Elsevier},
  doi={10.1016/j.aop.2019.167998},
  url={https://doi.org/10.1016/j.aop.2019.167998}
}

@article{motta2020determining,
  title={Determining eigenstates and thermal states on a quantum computer using quantum imaginary time evolution},
  author={Motta, Mario and Sun, Chong and Tan, Adrian TK and O’Rourke, Matthew J and Ye, Erika and Minnich, Austin J and Brand{\~a}o, Fernando GSL and Chan, Garnet Kin-Lic},
  journal={Nature Physics},
  volume={16},
  number={2},
  pages={205--210},
  year={2020},
  publisher={Nature Publishing Group UK London},
  doi={10.1038/s41567-019-0704-4},
  url={https://doi.org/10.1038/s41567-019-0704-4}
}

@article{mcardle2019variational,
  title={Variational ansatz-based quantum simulation of imaginary time evolution},
  author={McArdle, Sam and Jones, Tyson and Endo, Suguru and Li, Ying and Benjamin, Simon C and Yuan, Xiao},
  journal={npj Quantum Information},
  volume={5},
  number={1},
  pages={75},
  year={2019},
  publisher={Nature Publishing Group UK London},
  doi={10.1038/s41534-019-0187-2},
  url={https://doi.org/10.1038/s41534-019-0187-2}
}

@article{wilde2026scalably,
  title={Scalably learning one-dimensional quantum many-body Hamiltonians from dynamical data},
  author={Wilde, F and Kshetrimayum, A and Roth, I and Hangleiter, D and Sweke, R and Eisert, J},
  journal={Quantum Science and Technology},
  volume={11},
  number={3},
  pages={035002},
  year={2026},
  publisher={IOP Publishing},
  doi={10.1088/2058-9565/ae6fe3},
  url={https://doi.org/10.1088/2058-9565/ae6fe3}
}

@article{eisert2020quantum,
  title={Quantum certification and benchmarking},
  author={Eisert, Jens and Hangleiter, Dominik and Walk, Nathan and Roth, Ingo and Markham, Damian and Parekh, Rhea and Chabaud, Ulysse and Kashefi, Elham},
  journal={Nature Reviews Physics},
  volume={2},
  number={7},
  pages={382--390},
  year={2020},
  publisher={Nature Publishing Group UK London},
  doi={10.1038/s42254-020-0186-4},
  url={https://doi.org/10.1038/s42254-020-0186-4}
}

@article{kokail2021entanglement,
  title={Entanglement Hamiltonian tomography in quantum simulation},
  author={Kokail, Christian and van Bijnen, Rick and Elben, Andreas and Vermersch, Beno{\^\i}t and Zoller, Peter},
  journal={Nature Physics},
  volume={17},
  number={8},
  pages={936--942},
  year={2021},
  publisher={Nature Publishing Group UK London},
  doi={10.1038/s41567-021-01260-w},
  url={https://doi.org/10.1038/s41567-021-01260-w}
}

@article{lieb1972finite,
  title={The finite group velocity of quantum spin systems},
  author={Lieb, Elliott H and Robinson, Derek W},
  journal={Communications in Mathematical Physics},
  volume={28},
  number={3},
  pages={251--257},
  year={1972},
  publisher={Springer},
  doi={10.1007/BF01645779},
  url={https://doi.org/10.1007/BF01645779}
}

@article{cheneau2012light,
  title={Light-cone-like spreading of correlations in a quantum many-body system},
  author={Cheneau, Marc and Barmettler, Peter and Poletti, Dario and Endres, Manuel and Schau{\ss}, Peter and Fukuhara, Takeshi and Gross, Christian and Bloch, Immanuel and Kollath, Corinna and Kuhr, Stefan},
  journal={Nature},
  volume={481},
  number={7382},
  pages={484--487},
  year={2012},
  publisher={Nature Publishing Group UK London},
  doi={10.1038/nature10748},
  url={https://doi.org/10.1038/nature10748}
}

@article{proctor2020detecting,
  title={Detecting and tracking drift in quantum information processors},
  author={Proctor, Timothy and Revelle, Melissa and Nielsen, Erik and Rudinger, Kenneth and Lobser, Daniel and Maunz, Peter and Blume-Kohout, Robin and Young, Kevin},
  journal={Nature Communications},
  volume={11},
  number={1},
  pages={5396},
  year={2020},
  publisher={Nature Publishing Group UK London},
  doi={10.1038/s41467-020-19074-4},
  url={https://doi.org/10.1038/s41467-020-19074-4}
}

@article{kelly2018physical,
  title={Physical qubit calibration on a directed acyclic graph},
  author={Kelly, Julian and O'Malley, Peter and Neeley, Matthew and Neven, Hartmut and Martinis, John M},
  journal={arXiv preprint},
  year={2018},
  doi={10.48550/arXiv.1803.03226},
  url={https://doi.org/10.48550/arXiv.1803.03226}
}

@article{bairey2020learning,
  title={Learning the dynamics of open quantum systems from their steady states},
  author={Bairey, Eyal and Guo, Chu and Poletti, Dario and Lindner, Netanel H and Arad, Itai},
  journal={New Journal of Physics},
  volume={22},
  number={3},
  pages={032001},
  year={2020},
  publisher={IOP Publishing},
  doi={10.1088/1367-2630/ab73cd},
  url={https://doi.org/10.1088/1367-2630/ab73cd}
}

@article{samach2022lindblad,
  title={Lindblad tomography of a superconducting quantum processor},
  author={Samach, Gabriel O and Greene, Ami and Borregaard, Johannes and Christandl, Matthias and Barreto, Joseph and Kim, David K and McNally, Christopher M and Melville, Alexander and Niedzielski, Bethany M and Sung, Youngkyu and others},
  journal={Physical Review Applied},
  volume={18},
  number={6},
  pages={064056},
  year={2022},
  publisher={APS},
  doi={10.1103/PhysRevApplied.18.064056},
  url={https://doi.org/10.1103/PhysRevApplied.18.064056}
}

@article{ha2018world,
  title={World models},
  author={Ha, David and Schmidhuber, J{\"u}rgen},
  journal={arXiv preprint},
  year={2018},
  doi={10.48550/arXiv.1803.10122},
  url={https://doi.org/10.48550/arXiv.1803.10122}
}

@article{hafner2025mastering,
  title={Mastering diverse control tasks through world models},
  author={Hafner, Danijar and Pasukonis, Jurgis and Ba, Jimmy and Lillicrap, Timothy},
  journal={Nature},
  volume={640},
  number={8059},
  pages={647--653},
  year={2025},
  publisher={Nature Publishing Group UK London},
  doi={10.1038/s41586-025-08744-2},
  url={https://doi.org/10.1038/s41586-025-08744-2}
}

@article{assran2025v,
  title={V-jepa 2: Self-supervised video models enable understanding, prediction and planning},
  author={Assran, Mido and Bardes, Adrien and Fan, David and Garrido, Quentin and Howes, Russell and Komeili, Mojtaba and Muckley, Matthew and Rizvi, Ammar and Roberts, Claire and Sinha, Koustuv and Zholus, Artem and others},
  journal={arXiv preprint},
  year={2025},
  doi={10.48550/arXiv.2506.09985},
  url={https://doi.org/10.48550/arXiv.2506.09985}
}

@article{johnson2026hamiltonian,
  title={Hamiltonian Learning at Scale},
  author={Johnson, Nathan and Dumitrescu, Eugene},
  journal={arXiv preprint},
  year={2026},
  doi={10.48550/arXiv.2608.23707},
  url={https://doi.org/10.48550/arXiv.2608.23707}
}

@article{google2025constructive,
  title={Observation of constructive interference at the edge of quantum ergodicity},
  author={{Google Quantum AI and Collaborators}},
  journal={Nature},
  volume={646},
  pages={825--830},
  year={2025},
  doi={10.1038/s41586-025-09526-6}
}

@article{alam2025fermionic,
  title={Fermionic dynamics on a trapped-ion quantum computer beyond exact classical simulation},
  author={Alam, Faisal and Bosse, Jan Lukas and {\v{C}}epait{\.e}, Ieva and others},
  journal={arXiv preprint},
  year={2025},
  doi={10.48550/arXiv.2510.26300},
  url={https://doi.org/10.48550/arXiv.2510.26300}
}

@article{barron2026observable,
  title={Observable estimation in the absence of classical verification},
  author={Barron, Samantha V. and Mitchell, Bradley and Tripathi, Vinay and others},
  journal={arXiv preprint},
  year={2026},
  doi={10.48550/arXiv.2607.25998},
  url={https://doi.org/10.48550/arXiv.2607.25998}
}

@article{haghshenas2026digital,
  title={Digital quantum magnetism on a trapped-ion quantum computer},
  author={Haghshenas, Reza and Chertkov, Eli and Mills, Michael and Kadow, Wilhelm and Lin, S-H and Chen, Yi-Hsiang and Cade, Chris and Niesen, Ido and Begu{\v{s}}i{\'c}, Tomislav and Rudolph, Manuel S and others},
  journal={Nature},
  volume={653}, 
  pages={56--62},
  year={2026},
  publisher={Nature Publishing Group UK, London},
  doi={10.1038/s41586-026-10445-3},
  url={https://doi.org/10.1038/s41586-026-10445-3}
}

@inproceedings{bakshi2024learning,
  title={Learning quantum {H}amiltonians at any temperature in polynomial time},
  author={Bakshi, Ainesh and Liu, Allen and Moitra, Ankur and Tang, Ewin},
  booktitle={Proceedings of the 56th Annual ACM Symposium on Theory of Computing (STOC)},
  publisher={ACM},
  pages={1470--1477},
  year={2024},
  doi={10.1145/3618260.3649619}
}

@article{hu2025ansatz,
  title={Ansatz-free {H}amiltonian learning with {H}eisenberg-limited scaling},
  author={Hu, Hong-Ye and Ma, Muzhou and Gong, Weiyuan and Ye, Qi and Tong, Yu and Flammia, Steven T. and Yelin, Susanne F.},
  journal={PRX Quantum},
  volume={6},
  pages={040315},
  year={2025},
  doi={10.1103/j7b8-pb77}
}

@article{wang2017experimental,
  title={Experimental quantum {H}amiltonian learning},
  author={Wang, Jianwei and Paesani, Stefano and Santagati, Raffaele and Knauer, Sebastian and Gentile, Antonio A. and Wiebe, Nathan and Petruzzella, Maurangelo and O'Brien, Jeremy L. and Rarity, John G. and Laing, Anthony and Thompson, Mark G.},
  journal={Nature Physics},
  volume={13},
  pages={551--555},
  year={2017},
  doi={10.1038/nphys4074}
}

@article{gentile2021learning,
  title={Learning models of quantum systems from experiments},
  author={Gentile, Antonio A. and Flynn, Brian and Knauer, Sebastian and Wiebe, Nathan and Paesani, Stefano and Granade, Christopher E. and Rarity, John G. and Santagati, Raffaele and Laing, Anthony},
  journal={Nature Physics},
  volume={17},
  pages={837--843},
  year={2021},
  doi={10.1038/s41567-021-01201-7}
}

@inproceedings{herde2024poseidon,
  title={Poseidon: Efficient foundation models for {PDEs}},
  author={Herde, Maximilian and Raoni{\'c}, Bogdan and Rohner, Tobias and K{\"a}ppeli, Roger and Molinaro, Roberto and de B{\'e}zenac, Emmanuel and Mishra, Siddhartha},
  booktitle={Advances in Neural Information Processing Systems (NeurIPS)},
  publisher={Curran Associates, Inc.},
  volume={37},
  pages={72525--72624},
  year={2024},
  doi={10.52202/079017-2311},
  url={https://doi.org/10.52202/079017-2311}
}

@article{heim2021quantum,
  title={Quantum model-discovery},
  author={Heim, Niklas and Ghosh, Atiyo and Kyriienko, Oleksandr and Elfving, Vincent E.},
  journal={arXiv preprint},
  year={2021},
  doi={10.48550/arXiv.2111.06376},
  url={https://doi.org/10.48550/arXiv.2111.06376}
}

@article{mahassen2026quantum,
  title={Quantum {F}okker-{P}lanck solvers as trainable generative models},
  author={Mahassen, Tariq and Jastrzebski, Marcin and Kyriienko, Oleksandr},
  journal={arXiv preprint},
  year={2026},
  doi={10.48550/arXiv.2610.31330},
  url={https://doi.org/10.48550/arXiv.2610.31330}
}

@article{caro2023out,
  title={Out-of-distribution generalization for learning quantum dynamics},
  author={Caro, Matthias C. and Huang, Hsin-Yuan and Ezzell, Nicholas and Gibbs, Joe and Sornborger, Andrew T. and Cincio, Lukasz and Coles, Patrick J. and Holmes, Zo{\"e}},
  journal={Nature Communications},
  volume={14},
  pages={3751},
  year={2023},
  doi={10.1038/s41467-023-39381-w}
}

@article{azizzadenesheli2024neural,
  title={Neural operators for accelerating scientific simulations and design},
  author={Azizzadenesheli, Kamyar and Kovachki, Nikola and Li, Zongyi and Liu-Schiaffini, Miguel and Kossaifi, Jean and Anandkumar, Anima},
  journal={Nature Reviews Physics},
  volume={6},
  pages={320--328},
  year={2024},
  doi={10.1038/s42254-024-00712-5}
}

\end{document}